\documentstyle[aps,prb,eqsecnum,preprint,epsfig]{revtex}

\tightenlines

\begin{document}
\draft


\title{Instantaneous Pair Theory for High--Frequency Vibrational Energy
Relaxation in Fluids\\[2in]}

\author{Ross E. Larsen and Richard M. Stratt}

\address{Department of Chemistry, Brown University, Providence, RI
02912} \date{29 September, 1998}

\maketitle

\vspace*{4.25in}


\begin{abstract}

Notwithstanding the long and distinguished history of studies of
vibrational energy relaxation, exactly how it is that high frequency
vibrations manage to relax in a liquid remains somewhat of a mystery.
Both experimental and theoretical approaches seem to say that there is
a natural frequency range associated with intermolecular motion in
liquids, typically spanning no more than a few hundred $cm^{-1}$.
Landau--Teller--like theories explain rather easily how a solvent can
absorb any vibrational energy within this ``band'', but how is it that
molecules can rid themselves of superfluous vibrational energies
significantly in excess of these values?  In this paper we develop a
theory for such processes based on the idea that the crucial liquid
motions are those that most rapidly modulate the force on the vibrating
coordinate --- and that by far the most important of these motions are
those involving what we have called the mutual nearest neighbors of the
vibrating solute.  Specifically, we suggest that whenever there is a
single solvent molecule sufficiently close to the solute that the
solvent and solute are each other's nearest neighbors, then the
instantaneous scattering dynamics of the solute--solvent pair alone
suffices to explain the high frequency relaxation.  This highly reduced
version of the dynamics has implications for some of the previous
theoretical formulations of this problem.  Previous
instantaneous--normal--mode theories allowed us to understand the
origin of a band of liquid frequencies, and even had some success in
predicting relaxation within this band, but lacking a sensible picture
of the effects of liquid anharmonicity on dynamics, were completely
unable to treat higher frequency relaxation.  When
instantaneous--normal--mode dynamics is used to evaluate the
instantaneous pair theory, though, we end up with a multiphonon picture
of the relaxation which is in excellent agreement with the exact
high-frequency dynamics --- suggesting that the critical anharmonicity
behind the relaxation is not in the complex, underlying liquid
dynamics, but in the relatively easy--to--understand nonlinear
solute--solvent coupling.  There are implications, as well, for the
Independent-Binary-Collision (IBC) theory of vibrational relaxation in
liquids.  The success of the instantaneous--pair approach certainly
provides a measure of justification for the IBC model's focus on
few--body dynamics.  However, the pair theory neither needs nor
supports the basic IBC factoring of relaxation rates into many--body
and few--body dynamical components --- into collision rates and
relaxation rates per collision.  Rather, our results favor taking an
instantaneous perspective:  the relaxation rate is indeed exercise in
few--body dynamics, but a different exercise for each instantaneous
liquid configuration.  The many--body features therefore appear only in
the guise of a purely equilibrium problem, that of finding the
liklihood of particularly effective solvent arrangements around the
solute.  All of these results are tested numerically on model diatomic
solutes dissolved in atomic fluids (including the experimentally and
theoretically interesting case of I$_2$ dissolved in Xe).  The
instantaneous pair theory leads to results in quantitative agreement
with those obtained from far more laborious exact molecular dynamics
simulations.

\end{abstract}

\newpage

\section{Introduction}\label{sec1}

The most crucial step in a solution--phase chemical reaction is almost
always a thermal activation; improbable as it may be, the solvent must
find some avenue for concentrating enough of its kinetic energy within
the vibrations of the reactants to allow the reaction to
proceed.~\cite{hynes85,tokmakoff94}  A similar challenge can be posed
later on in the process; the ability of the resulting products to
return any excess vibrational energy back to the solvent is often
critical to the outcome of the reaction.~\cite{hynes85,pollak96} Beyond
these issues however, there is a conceptual importance to understanding
such vibrational energy transfers that goes to the heart of how one
thinks about the dynamical role of a solvent.  If the entire
solvent-solute system is regarded as a ``super molecule'' (as might be
appropriate for solid-state vibrational relaxation), then the question
is how a V--V relaxation occurs --- that is, which vibrations of the
system as a whole (which phonons) can serve as repositories for the
solute's vibrational
energy.~\cite{dlott89,egorov95_1,egorov95_2,egorov97}  If, on the other
hand, we consider the solute to be buffeted by occasional collisions
with individual solvent molecules, a very natural perspective for the
gas phase, what we need to understand is a V--T
process.~\cite{vt-process}  So how is it that we should think about
vibrational energy relaxation in
liquids?~\cite{owrutsky94,harris90,chesnoy88,berne90,tuckerman93,whitnell90,bruehl93,rey96,egorov96,adelman93,miller94,tokmakoff95,gnanakaran96,hamm97}

A formal answer to this question arises quite simply from a Fermi's
Golden Rule evaluation of the relaxation rate.~\cite{oxtoby81}  Within
the apparently quite broad reach of low--order time--dependent
perturbation theory, the rate at which vibrational energy relaxes from
a given solute mode, $T_1^{-1}$, is proportional to the Fourier
component of the autocorrelation function of the force on that mode at
the mode's own frequency, $\omega_0$.~\cite{renormed-w}  For a diatomic
molecule of reduced mass $\mu$, for example, the classical vibrational
energy relaxation rate is predicted to obey the Landau--Teller
formula,~\cite{whitnell90} \begin{eqnarray} {1 \over T_1}\ =\ {1 \over
\mu} {\hat \eta}(\omega_0)\ , \label{l-t-form} \end{eqnarray} where
\begin{eqnarray} {\hat \eta}(\omega_0)\ =\ \int_0^\infty dt\ \eta(t)
\cos{\omega_0 t} \label {define-fric} \\ \cr \eta(t)\ =\ {1 \over k_B
T} \left< \delta F(t)\ \delta F(0) \right>\ .  \label{define-fcorr}
\end{eqnarray}  Here $k_B$ is Boltzmann's constant, $T$ is the
temperature, and by $\delta F(t) = F(t) - F$, the fluctuating force on
the mode, we mean the hypothetical classical time evolution of the
solvent that would occur were the mode itself held motionless.

Leaving aside such issues as whether a purely classical treatment of
the solvent can adequately represent the influence on what is
frequently a rather quantum mechanical vibration,~\cite{quantum} it
seems clear that this approach can usually be counted on to give a
numerically accurate rendition of classical vibrational energy
relaxation.~\cite{tuckerman93}  More than that, though, it provides a
conceptually suggestive rendition because it fits so nicely into the
broader context of relaxation in liquids.  Within a Generalized
Langevin representation of the problem, the function $\eta(t)$ can be
interpreted as an approximation to the solvent-induced friction felt by
a vibrating coordinate --- and since the exact friction can sometimes
be determined by simulation, it has been possible both to check and to
understand the formalism's accuracy.~\cite{berne90}

Still, the fact that numerical success does not necessarily translate
into conceptual insight becomes clear once we try to get at the central
V--V vs. V--T dichotomy.  It has long been appreciated that any
frequency--domain friction ${\hat \eta}(\omega)$ can be regarded as
having its origin in the dynamics of a dense set of harmonic
oscillators, oscillators whose frequencies span the frequency range of
the friction and which are coupled linearly to the coordinate of
interest.~\cite{deutch71,levine88}  The fact that liquid frictions
interpreted in this fashion typically have a spectrum spanning a few
hundred $cm^{-1}$ could thus be taken as evidence that there is indeed
something resembling a phonon band of this width in liquids.  It is
then but a short step to suggest that any vibrational relaxation of
modes with frequencies lying within this band ought to be considered as
occuring via a resonant V--V transfer.

This argument is actually rather weak as it stands.  These harmonic
oscillators could, in fact, be little more than mathematical constructs
lacking any microscopic, molecular foundation.  What gives the argument
a bit more substance is that recent work has shown that it is possible
to associate these oscillators with the instantaneous normal modes of
the solutions, oscillators which really are specific, well--defined
molecular
motions.~\cite{goodyear96_1,goodyear96_2,goodyear97,larsen97,ladanyi98,kalbfleisch98,schvan95,gershinsky94}
Indeed, perhaps it is fair to say that the basic success of such
instantaneous--normal--mode (INM) approaches in predicting most of the
essential behavior of the vibrational friction lends a certain amount
of credibility to what would otherwise be fanciful attempts to take
liquid phonon pictures literally in theories of vibrational
relaxation.~\cite{multi-phon}

There is, however, a major problem and it is this problem that forms
the topic of this paper.  Once the vibration to be relaxed lies outside
the putative band of the liquid, any theory with nothing but linear
forces and harmonic oscillators will predict that energy relaxation is
completely forbidden.~\cite{intra-relax}  Yet, what we might call the
high frequency modes of solutes (those with frequencies higher than the
Debye frequency in the solid--state parlance) really do relax, albeit
very slowly.~\cite{owrutsky94,egorov96,nitzan97}  It is worth noting
that the Landau--Teller formula itself has no difficulty with this
phenomenon; it simply ascribes the relaxation to the presence of small
but finite high--frequency Fourier components of the force
autocorrelation function.~\cite{whitnell90}  The conceptual difficulty
is that we have independent evidence about the intrinsic bandwidth of
intermolecular vibrations in a liquid, both from theoretical
predictions, such as instantaneous--normal--mode
analysis~\cite{stratt95,ladanyi96,moore94} and Fourier representations
of simulated velocity autocorrelation functions~\cite{rahman76} and
from experimental results, such as those from far infrared, Rayleigh
and Optical Kerr effect spectra.~\cite{mol-liq,mcmorrow91}  None of
these estimates give any evidence that the band for simple liquids
extends significantly beyond a few hundred $cm^{-1}$.  A much more
likely scenario than one that would require every simple liquid to have
the kinds of 500--2000 $cm^{-1}$ intermolecular vibrations necessary
for simple V--V processes is that high frequency vibrational energy
relaxation is a strongly anharmonic event --- and it is this
anharmonicity we need to understand.

Having made this point, though, there is an interesting distinction we
can make.  The relaxation of impurity molecules in crystals can often
be modeled as a multiphonon process, one in which the essential
harmonicity of the vibrations (and hence the very idea of phonons) is
respected, but in which the nonlinearity of the coupling forces allows
for V-V transfer to overtones of the crystal's intrinsic
frequencies.~\cite{egorov95_1,egorov95_2,egorov97,nitzan74,nitzan75} In
systems as inherently anharmonic as liquids, however, we could easily
imagine that the basic idea of independent harmonic excitations could
be the first casualty.~\cite{passino97,okumura97,david98}  Thus,
without further work, it is far from clear which of these two kinds of
anharmonicity --- the nonlinearity of the forces being driven by the
liquid's dynamics or that of the actual dynamics itself --- ought to be
more critical to high--frequency relaxation in liquids.  In particular,
we are still left with the question of whether the V--V paradigm for
resonant energy transfer is going to be any better than a V--T based
perspective would be.

Complicating this issue is the fact that there is some evidence on the
side of molecular translation being the primary sink for vibrational
energy.  That is, there is some support for a collisional model for
vibrational relaxation.  While it is admittedly far from intuitive how
discrete, uncorrelated collisions could ever occur in a dense a medium
as a liquid, the so--called Isolated--Binary Collision (IBC)
model~\cite{davis72,oxtoby77} has been remarkably successful in
rationalizing the thermodynamic state dependence of vibrational
population relaxation rates based on just such a
picture.~\cite{chesnoy86,paige90,dardi88}  The notion behind this model
is that these rates should factor into a purely few--body scattering
piece specifying the rate of energy transfer per collision, and a
liquid--state piece quantifying the rate at which the solvent collides
with the solute.  The fact that the concept (and thus the rate) of
collisions is ill--defined in liquids is certainly
vexing,~\cite{fixman61,dardi88,simpson96} but it is hardly out of the
question that some other, more sophisticated analysis of the collision
concept might find that the V--T motif is still a dominant mechanism
for relaxation.

The approach we shall take in this paper to getting at these issues is
to begin with the standard perturbative (Landau--Teller) expression in
terms of the force autocorrelation function, Eqs.
(\ref{l-t-form})--(\ref{define-fcorr}), but then ask from precisely
where we should expect the major high frequency components of the
correlation function to arise.  For any instantaneous configuration of
the liquid, this analysis will immediately lead us to look at the
short--time scattering dynamics of the solute and a single crucial
solvent partner.  Previous work on vibrational relaxation has shown
that whenever there is a solvent molecule that is nearer to the solute
than any other solvent molecule is to either member of the
solute--solvent pair --- that is, whenever there is a
mutual--nearest--neighbor solute--solvent pair --- then the
high-frequency solvent effects are dominated by the pair, at least
within the liquid's band.~\cite{larsen97}  We will show how this
observation can be exploited more generally, transcending the
limitations to dynamics within the band.  Not surprisingly, our
theoretical prediction for the relaxation rate will then separate into
a straightforward few--body dynamical calculation and a residual
many--body problem, just as with IBC theory.  However, unlike the IBC
approach, the liquid--state portion of our theory will turn out to rest
on a simple, well--defined, purely equilibrium question.  Moreover, our
development will make it clear how we can make connections between this
V--T theory and the harmonic instantaneous dynamics of
instantaneous--normal--modes, and thereby connect with the V--V energy
transfer ideas which seem to work so well within the solvent band.

The remainder of the paper will be organized as follows:  Section
\ref{sec2} will present our basic theory, an expression for the full
force autocorrelation function of the system in terms of a correlation
function for single degree of freedom, one whose equations of motion we
proceed to derive.  Section III then explores the connection to the
instantaneous--normal--mode perspective, showing how we can view the
new theory as a natural generalization of INM ideas.  Section IV
provides some numerical examples of how and how well the various
approaches work, and we conclude in Sec. V with some general comments
on what we are now in a position to say about the questions posed in
this Introduction.

\section{Instantaneous Pair Theory for the Vibrational Friction}
\label{sec2}

\subsection{Solute--Solvent Coupling versus Solute--Solvent Dynamics}
\label{coup-dyn-sec}

While the central quantity of interest in vibrational relaxation is the
force autocorrelation function of Eq. (\ref{define-fcorr}), it is
convenient to look instead at a mathematically equivalent function, the
``force--velocity'' autocorrelation function,~\cite{stratt94,steele87}
\begin{eqnarray} G_F(t)\ =\ \left< {\dot F}(t)\ {\dot F}(0)
\right>\ =\ -(k_B T)\ {\ddot \eta}(t)\ , \label{g-def} \end{eqnarray}
which lets us write the frequency-domain vibrational friction we need
in terms of the cosine transform of the new
function~\cite{goodyear96_1,goodyear96_2,delta-fric} \begin{eqnarray}
{\hat \eta}(\omega)\ =\ {1 \over k_B T}\ \left( {1 \over \omega^2}
\right) \int_0^\infty dt\ G_F(t)\ \cos{\omega t}\ \equiv\ {1 \over k_B
T}\ \left( {{\hat G}_F(\omega) \over \omega^2} \right) \ .
\label{fric-G-omega} \end{eqnarray}  The main reason we bother with
this conversion is that this alternative version will suggest a natural
division within our problem.  As one might have expected, there will
always be contributions from the relative solute--solvent velocities
--- the intrinsic dynamics of the problem --- but there will be also be
separate features dependent directly on the solute--solvent distances
--- and it is these that will be most pertinent to the solute--solvent
coupling.~\cite{ladanyi98}  Indeed, one of the key aspects of our
subsequent development will be the realization that these two kinds of
information actually play substantially different roles in determining
the high--frequency behavior.

We can begin to see how this division occurs by looking at the problem
of evaluating a general correlation function of the form of Eq.
(\ref{g-def}) when $F$ is any specifically solute-centered quantity.
That is, suppose we look at a correlation function of the form of Eq.
(\ref{g-def}) with the function $F$ given by a sum of solute-solvent
pair ``potentials'', \begin{eqnarray} F\ =\ \sum_{j\ \in\ solvent}
w(r_{0j})\ , \end{eqnarray} with the tagged solute labeled as particle
0, the remaining solvent denoted by $j\ =\ 1,\ldots ,\ N$ and the
interparticle distances given by $r_{0j} \equiv |{\bf r_0} - {\bf
r_j}|$, so that \begin{eqnarray} {\dot F}\ =\ \sum_{j\ \in\ solvent}
w^\prime(r_{0j})\ {\dot r}_{0j}\ . \label{f-dot-def} \end{eqnarray}
This general function $w(r)$, something we have referred to elsewhere
as the ``spectroscopic probe potential'',~\cite{ladanyi98} could very
well be the force on some portion of the solute (as it is in
vibrational relaxation), but for the purposes of this argument it could
also be the change in the solute-solvent potential energies when the
solute is promoted from one electronic state to another (the situation
in time-dependent fluorescence studies of solvation).~\cite{ladanyi96}

If we substitute Eq. (\ref{f-dot-def}) into Eq. (\ref{g-def}), we find
that the velocity form of the correlation function consists of two
kinds of terms, \begin{eqnarray} G_F(t) &=& \left< \sum_j
w^\prime[r_{0j}(t)]\ w^\prime[r_{0j}(0)]\ {\dot r}_{0j}(t)\ {\dot
r}_{0j}(0) \right> \cr\cr &+& \left< \sum_{j, k\ (j \neq k)}
w^\prime[r_{0j}(t)]\ w^\prime[r_{0k}(0)]\ {\dot r}_{0j}(t)\ {\dot
r}_{0k}(0)\right>\ , \label{g-f-explicit} \end{eqnarray} a binary
piece, the first term, which considers only the dynamics caused by the
motion of a single solvent at a time, and a ternary piece, the second
term, which involves the simultaneous motion of two different
solvents.

These two parts will not contribute equally, for reasons we shall come
to presently, but the essential ingredients of the correlation function
are already visible.  When we evaluate any of the terms at time $t =
0$, \begin{eqnarray} \left< w^\prime[r_{0j}(0)]\
w^\prime[r_{0k}(0)]\ {\dot r}_{0j}(0)\ {\dot r}_{0k}(0) \right> &=&
\left< w^\prime[r_{0j}(0)]\ w^\prime[r_{0k}(0)]\right>\ \left< {\dot
r}_{0j}(0)\ {\dot r}_{0k}(0) \right> \cr\cr &=& \delta_{jk} \left<
w^\prime[r_{0j}(0)]\ w^\prime[r_{0j}(0)]\right>\ \left< {\dot
r}_{0j}(0)\ {\dot r}_{0j}(0) \right>\ , \label{g-at-time0}
\end{eqnarray} we notice that the contribution factors rigorously into
a static part involving the coupling, $w^\prime(r)$, and a dynamical
part involving the velocity, ${\dot r}$.  This separation suggests,
though, that this same conceptual distinction between coupling and
dynamics ought to continue to operate at short times --- and therefore
at high frequencies --- despite the fact that the expression no longer
factors.  But, once we make such a distinction, we can simplify our
expressions dramatically.  For one thing, at the highest frequencies we
should expect coupling to be far more influential than dynamics in
determining the correlation function.  With typical intermolecular
potentials and forces, the sharply varying repulsive parts allow for
tremendous variation in the coupling (and therefore enormous values of
$w^\prime(r))$ just from small changes in internuclear distances.  The
velocities, by contrast, never become all that large in magnitude,
guided as they are by the equilibrium Boltzmann distribution.

We can make this same point a bit more quantitatively.  If we assume
that the $t = 0$ factorization holds, at least roughly, then each term
in $G_F(t)$ can be written \newpage \begin{eqnarray}
T_{jk}(t)\ &=&\ \left< w^\prime[r_{0j}(t)]\ w^\prime[r_{0k}(0)]\ {\dot
r}_{0j}(t)\ {\dot r}_{0k}(0) \right>\ \cr\cr &\approx&\ \left<
w^\prime[r_{0j}(t)]\ w^\prime[r_{0k}(0)] \right>\ \left< {\dot
r}_{0j}(t)\ {\dot r}_{0k}(0) \right>\ ,\nonumber \end{eqnarray} which
becomes, on cosine transforming the resulting product, \begin{eqnarray}
{\hat T}_{jk}\ \approx\ {\delta_{jk} \over \pi}\ \int_0^\infty
d\omega^\prime\ {\hat V}(\omega^\prime) \left[ {\hat C}(\omega -
\omega^\prime) + {\hat C}(\omega + \omega^\prime) \right] \ , \nonumber
\end{eqnarray} where ${\hat V}(\omega)$ and ${\hat C}(\omega)$ are the
cosine transforms of the single--solvent velocity and coupling
correlation function, respectively.  What then determines the high
frequency behavior?  The intrinsic frequencies of the liquid are
largely those of ${\hat V}(\omega^\prime)$,~\cite{rahman76} which cuts
off at frequencies $\omega^\prime$ well below the $\omega$ we are
interested in.  But if $\omega^\prime \ll \omega$, then ${\hat
C}(\omega - \omega^\prime)\ \approx {\hat C}(\omega)$, so
\begin{eqnarray} T_{jk}(\omega)\ \approx\ \delta_{jk} {2 \over \pi}
\left[ \int_0^\infty d \omega^\prime\ {\hat V}(\omega^\prime) \right]
{\hat C}(\omega)\ .  \end{eqnarray}  Hence, we see again that the
high--frequency dependence of our friction spectrum results almost
entirely from the high-frequency behavior of the coupling.

This argument can be taken still further.  The dynamic range of the
coupling is so large that we might expect it to be dominated by the
{\it very} largest contributions.  No more than a few of the closest
solvents should therefore play much of a role.  Indeed, we might
venture that only the single largest contribution, from the single most
strongly coupled solvent is important,~\cite{larsen97,ladanyi98}
\begin{eqnarray} G_F(t)\ =\ \left<
w^\prime[r_{0j}(t)]\ w^\prime[r_{0j}(0)]\ {\dot r}_{0j}(t)\ {\dot
r}_{0j}(0) \right>_{j=maximally\ coupled}\ .  \end{eqnarray}  What we
suggest here is that this latter idea is virtually exact; that the
high-frequency dynamics of a solute-centered correlation function stems
almost entirely from the binary dynamics of a single solute-solvent
pair --- and all we need do is compute the nonlinear, anharmonic
dynamics of this pair for each instantaneous liquid configuration.

There are actually a number of kinds of evidence supporting our
analysis.  An immediate consequence of our argument is that the binary
terms in Eq. (\ref{g-f-explicit}) ought to be significantly larger than
the ternary terms at short times.  Indeed, exact molecular dynamics
simulations have established the overwhelming predominance of the
binary contributions to the $G_F(t)$ vibrational relaxation correlation
functions.~\cite{ladanyi98}  Interestingly, one finds the predicted
strong preferences for the (presumably) less collective binary part
even when the relaxation takes place in dipolar solvents, liquids whose
intermolecular forces are sufficiently long--ranged that the friction
$\eta(t)$ itself ends up with major contributions from ternary
terms.~\cite{whitnell90,ladanyi98} The reasons for this difference
between binary and ternary are clear from a short--time perspective.
At equilibrium ($t = 0$), the force--velocity version of the
vibrational friction is rigorously binary, because, as we can see from
Eq.  (\ref{g-at-time0}), the velocities of different solvent molecules
are rigorously independent.  The actual equilibrium friction, by
contrast, does not even start out as binary at $t = 0$; the equilibrium
structure of the liquid builds in correlations that have nothing to do
with the subsequent dynamical evolution.  But even away from $t = 0$,
the wide range of values possible for the couplings serves to explain
the observation.  The sheer unlikelihood of having two different
solvents so similar in their distance from the solute that they end up
making comparable contributions will also prevent much of a ternary
presence in the answer.

We can see somewhat more direct evidence of the two-particle character
of the essential dynamics by looking in the frequency domain with the
aid of INM theory.  INM theories cannot addresss any of the truly
high-frequency dynamics, but if we confine ourselves to the highest
frequencies inside the INM band, we invariably find that the
vibrational friction spectrum ($\rho_{vib}(\omega) = (2/\pi) {\hat
G}_F(\omega)$) is given quantitatively by the portion arising from the
motion of the solute and the single largest contributing
solvent.~\cite{larsen97,ladanyi98} Moreover, as with the molecular
dynamics studies, these findings seems to be completely independent of
how otherwise collective the solvent motion is.  The question we are
left with, therefore, is whether the nature of the relaxation really
does change qualitatively as we cross the band edge.  It does not seem
all that far fetched for this same two--particle scenario that works so
well just inside the band to apply at higher frequencies as well.

\subsection{The High-Frequency Parts of the Vibrational Relaxation
Correlation Function} \label{high-freq}

To make our subsequent development more concrete, suppose we specialize
to the familiar case of a diatomic solute dissolved in an atomic
liquid.  Taking the atoms of the diatomic to be labeled as 1 and 2
(with masses $m_1$ and $m_2$), and indexing the atoms of the solvent by
$j = 3,\ \ldots,\ N$, we find that if the potential energy of
interaction between the solute and the solvent can be written in terms
of site-site pair potentials \begin{eqnarray} V\ =\ \sum_{j > 2} \left[
u_1(r_{1j}) + u_2(r_{2j}) \right] \ , \label{v-diatomic} \end{eqnarray}
then the force acting on the diatomic's bond is given by
\begin{eqnarray} F_x\ =\ - {\partial V \over \partial x}\ =\ \mu_{12}
\sum_{j > 2} \left[ {1 \over m_1} u^\prime_1(r_{1j}) {\hat {\bf
r}}_{1j}\ -\ {1 \over m_2} u^\prime_2(r_{2j}) {\hat {\bf r}}_{2j}
\right] \cdot {\hat {\bf r}}_{12} \ . \label{bond-force} \end{eqnarray}
Here $\mu_{12} = m_1 m_2 / (m_1 + m_2)$ is the solute reduced mass and
${\hat {\bf r}} = {\bf r} / |{\bf r}|$.  The requisite time derivative
of the bond force is thus \begin{eqnarray} {\dot F}_x\ &=&\ \mu_{12}
\sum_{j > 2} \left[ {1 \over m_1} u^{\prime\prime}_1(r_{1j}) {\dot
r}_{1j} ( {\hat {\bf r}}_{1j} \cdot {\hat {\bf r}}_{12}) \ -\ {1 \over
m_2} u^{\prime\prime}_2(r_{2j}) {\dot r}_{2j} ( {\hat {\bf r}}_{2j}
\cdot {\hat {\bf r}}_{12}) \right] \cr\cr &+&\ \mu_{12} \sum_{j > 2}
\left[ {1 \over m_1} u^\prime_1(r_{1j}) {d \over dt} ( {\hat {\bf
r}}_{1j} \cdot {\hat {\bf r}}_{12})\ -\ {1 \over m_2}
u^\prime_2(r_{2j}) {d \over dt} ( {\hat {\bf r}}_{2j} \cdot {\hat {\bf
r}}_{12}) \right] \ .  \label{bond-force-v} \end{eqnarray}

Let us now consider what the highest frequency parts of the
force--velocity correlation function, Eq. (\ref{g-def}), would look
like for such a force.  Generalizing our arguments in the previous
section, we would expect that the dominant contribution would come (at
most) from the two solvent atoms, call them $a$ and $b$, nearest the
two solute sites 1 and 2 (respectively).~\cite{pairings}  Furthermore,
given the rapid variation of the site-site potentials $u_{1,2}(r)$ in
the region of interest, we might also expect the terms involving the
second derivative $u^{\prime\prime}(r)$ to be far more important than
those involving the first derivative $u^\prime(r)$.  Hence we should be
able to express the correlation function as just \newpage
\begin{eqnarray} G_F(t)\ &\approx&\ \left( {\mu_{12} \over m_1}
\right)^2 \left<
u^{\prime\prime}(r_{1a}(t))\ u^{\prime\prime}(r_{1a}(0))\
\cos{\theta_{1a}(t)}\ \cos{\theta_{1a}(0)}\ {\dot r}_{1a}(t)\ {\dot
r}_{1a}(0) \right> \cr\cr &+&\ \left( {\mu_{12} \over m_2} \right)^2
\left< u^{\prime\prime}(r_{2b}(t))\ u^{\prime\prime}(r_{2b}(0))\
\cos{\theta_{2b}(t)}\ \cos{\theta_{2b}(0)}\ {\dot r}_{2b}(t)\ {\dot
r}_{2b}(0) \right>\ ,\label{almost-there} \end{eqnarray} with
$\cos{\theta_{Aj}(t)} \equiv ({\bf {\hat r}_{Aj}} \cdot {\bf {\hat
r}_{12}})\ (A = 1, 2)$.  We could, in fact, make use of Eq.
(\ref{almost-there}) as it stands;  since it requires that we know no
more than the dynamics of 4 degrees of freedom, it already represents a
major simplification over the full many--body problem.  We can make one
last simplification, however.  The variation of the
$\cos{\theta_{Aj}(t)}$ angular factors is also guaranteed to be far
smaller than that of the $u^{\prime\prime}(r)$ factors at short times,
if only because of the small dynamic range of the cosine function ($-1
\leq \cos{\theta} \leq 1$).  For the purposes of evaluating
high--frequency behavior it therefore suffices to approximate the
cosine by its $t = 0$ value.  Our final high--frequency expression for
the vibrational relaxation correlation function thus reduces to a
remarkably simple form \begin{eqnarray} G_F(t)\ &\approx&\ \left(
{\mu_{12} \over m_1} \right)^2 \left<
u^{\prime\prime}(r_{1a}(t))\ u^{\prime\prime}(r_{1a}(0))\
\cos^2{\theta_{1a}(0)}\ {\dot r}_{1a}(t)\ {\dot r}_{1a}(0) \right>
\cr\cr &+&\ \left( {\mu_{12} \over m_2} \right)^2 \left<
u^{\prime\prime}(r_{2b}(t))\ u^{\prime\prime}(r_{2b}(0))\
\cos^2{\theta_{2b}(0)}\ {\dot r}_{2b}(t)\ {\dot r}_{2b}(0) \right>\ ,
\label{nn-g-theory} \end{eqnarray} relying only on the
one--dimensional, purely radial dynamics of two solvent molecules
moving with respect to the two solute sites.

\subsection{Mutual--Nearest--Neighbor Dynamics} \label{mnn-dyn-sec}

Of course, since we are in a liquid, the exact dynamics of even
this small a number of degrees of freedom is still intimately connected
with the dynamics of the remainder.  Were we unable to achieve any
reduction of this dynamics comparable to the simplification we found for
the correlation function, there would thus be little practical advantage to
our formalism.  Fortunately, the same kinds of analysis we have been
pursuing also reveal a wonderful dynamical separability:  the important
solvents end up being dynamically decoupled from the rest of the solvent as
well.

This basic notion of dynamical separability is illustrated by looking
at the time evolution of a mutual--nearest--neighbor pair of atoms in
an atomic liquid.~\cite{larsen97,david98}  The equations of motion for
any two atoms $a$ and $b$, with masses $m_a$ and $m_b$, are
\begin{eqnarray} m_a {\bf {\ddot r}}_a\ &=&\ u^\prime(r_{ab}) {\hat
{\bf r}}_{ab}\ -\ \sum_{j \neq a,b} u^\prime(r_{aj}) {\hat {\bf
r}}_{ja} \cr\cr m_b {\bf {\ddot r}}_b\ &=&\ - u^\prime(r_{ab}) {\hat
{\bf r}}_{ab}\ -\ \sum_{j \neq a,b} u^\prime(r_{bj}) {\hat {\bf
r}}_{jb} \ , \nonumber \end{eqnarray} provided all the atoms interact
by pair potentials $u(r)$.  But, if these two particular atoms form a
mutual--nearest--neighbor pair, then $u^\prime(r_{ab})$ will be
significantly larger than any of the other forces affecting the pair.
Hence, to within a high level of accuracy, the relative acceleration
will be given by \begin{eqnarray} {\bf {\ddot r}}_{ab}\ =\ {\bf {\ddot
r}}_b\ -\ {\bf {\ddot r}}_a\ \approx\ - {1 \over \mu}
u^{\prime}(r_{ab}) {\bf {\hat r}}_{ab}\ , \nonumber \end{eqnarray}
meaning that the scalar displacement $r_{ab}$ will satisfy the differential
equation \begin{eqnarray} {\ddot r}_{ab}\ =\ -{1 \over \mu}
u^\prime(r_{ab})\ +\ r_{ab} \Omega_{ab}^2\ , \label{pair-with-cent}
\end{eqnarray} with $\mu = m_a m_b / (m_a + m_b)$ the reduced mass of
the pair.  Here the angular velocity factor $\Omega_{ab}^2 \equiv -
{\bf {\hat r}}_{ab} \cdot {\bf {\ddot {\hat r}}}_{ab}$ and we have used
the vector identity \begin{eqnarray} {\bf {\hat r}}_{ab} \cdot {\bf
{\ddot r}}_{ab}\ =\ {\ddot r}_{ab}\ -\ r_{ab} \Omega_{ab}^2 \ .
\nonumber \end{eqnarray}

Equation (\ref{pair-with-cent}) is intriguing, but it is not as simple
as it may appear; the evolution of the centrifugal acceleration term
$r_{ab} \Omega_{ab}^2$ is controlled by other forces besides
$u^\prime(r_{ab})$ (and thus by the motion of other atoms besides $a$
and $b$).  However it is fairly easy to see that this term too is
almost always going to be negligible compared to the intra--pair
force.  We can estimate its size by using the equipartion theorem to
get the mean--square angular velocity~\cite{avging} $\Omega_{ab}^2 =
<\omega^2 > = (k_B T/\mu r_{ab}^2)$.  If we take the pair potential to
be of the Lennard--Jones form, $u(r) = 4 \epsilon [(\sigma/r)^{12} -
(\sigma/r)^6]$, and we then make the (very) conservative estimate that
a typical intra--pair distance $r_{ab} = \sigma$, we find that a
typical value of the centrifugal force is $(k_B T/\epsilon)
(\epsilon/\sigma)$, whereas a typical intermolecular force is $24
(\epsilon/\sigma)$.  Hence under any liquid--state conditions, indeed
at anything other than the most extreme supercritical conditions, we
can safely neglect such centrifugal contributions.~\cite{mnn-dists}
The dynamics of a mutual--nearest--neighbor pair of atoms is therefore
well described by a simple, purely radial, equation of motion:
\begin{eqnarray} {\ddot r}_{ab}\ =\ -{1 \over \mu} u^\prime(r_{ab})\ ,
\label{v-v-eqn} \end{eqnarray} even in a dense liquid.

This same argument can easily be adapted to the vibrational relaxation
example we considered in the last section.  If only one atom (say
number 1) of our diatomic solute forms a mutual nearest neighbor pair
with a solvent atom (solvent $a$), then the equations of motion for the
critical solvent atom and for ${\bf R}_{cm}$, the position of the
diatomic's center of mass, are given accurately by~\cite{c-m-force}
\begin{eqnarray} m_a {\bf {\ddot r}}_a\ &=&\ - u_1^\prime(r_{1a}) {\hat
{\bf r}}_{1a} \\ \cr M_{12} {\bf {\ddot
R}}_{cm}\ &=&\ u_1^\prime(r_{1a}) {\hat {\bf r}}_{1a} \label{c-m-eqn}
\end{eqnarray}  where $m_a$ is the mass of the solvent atom, and
$M_{12} = m_1 + m_2$ is the total mass of the solute.  For a rigid
diatomic, though, \begin{eqnarray} {\bf {\ddot r}}_1\ =\ {\bf {\ddot
R}}_{cm}\ -\ \left( {m_2 \over M_{12}} \right) r_{12} {\bf {\ddot {\hat
r}}}_{12} \nonumber \end{eqnarray} so that the analogue of Eq.
(\ref{pair-with-cent}) becomes \begin{eqnarray} {\ddot r}_{1a}\ =\ - {1
\over \mu_a} u_1^\prime(r_{1a})\ +\ r_{1a} \Omega_{1a}^2\ +\ \left(
{m_2 \over M_{12}} \right) r_{12} {\bf {\hat r}}_{1a} \cdot {\bf {\ddot
{\hat r}}}_{12} \ , \label{pair-dyn-with-rots} \end{eqnarray} with
$\mu_a = m_a M_{12} / (m_a + M_{12})$ the solute--solvent reduced
mass.

As we did before, we can now turn to the non--radial terms.  The
second, solute--solvent centrifugal, term in Eq.
(\ref{pair-dyn-with-rots}) can be discarded for precisely the same
reasons we just went through; on the average this term is far smaller
than the radial force $u^\prime_1(r_{1a})$.  The third term seems to
present a new case, inasmuch as it arises directly from the rotational
motion of the diatomic solute.  Yet, we can argue that it too should be
small.  Barring pathological situations, the interatomic separation and
the reduced mass for the diatomic should be comparable to those of a
solute--atom/solvent--atom mutual--near--neighbor pair.  We should
therefore expect similar results for the centrifugal terms
\begin{eqnarray} \Omega_{12}^2\ \equiv\ - {\bf {\hat r}}_{12} \cdot
{\bf {\ddot {\hat r}}}_{12} \ \sim\ \Omega_{1a}^2\ \equiv - {\bf {\hat
r}}_{1a} \cdot {\bf {\ddot {\hat r}}}_{1a} \ , \nonumber \end{eqnarray}
both of which are equal to angular velocity factors on the average
\begin{eqnarray} \left< \Omega_{12}^2 \right> = \left< {\bf {\dot {\hat
r}}}_{12} \cdot {\bf {\dot {\hat r}}}_{12}\right>\ ,\ \left<
\Omega_{1a}^2 \right> = \left< {\bf {\dot {\hat r}}}_{1a} \cdot {\bf
{\dot {\hat r}}}_{1a} \right> \nonumber \end{eqnarray} and should
therefore be small.  By the same token, the average rotational term $ <
- {\bf {\hat r}}_{1a} \cdot {\bf {\ddot {\hat r}}}_{12} >\ =\ < {\bf
{\dot {\hat r}}}_{1a} \cdot {\bf {\dot {\hat r}}}_{12} >$.  But by
Schwarz's inequality \begin{eqnarray} \left| {\bf {\dot {\hat r}}}_{1a}
\cdot {\bf {\dot {\hat r}}}_{12} \right|\ \leq\ \left( {\bf {\dot {\hat
r}}}_{12} \cdot {\bf {\dot {\hat r}}}_{12}\right)^{1/2}\ \left( {\bf
{\dot {\hat r}}}_{1a} \cdot {\bf {\dot {\hat r}}}_{1a} \right)^{1/2}
\ . \nonumber \end{eqnarray}  Hence, the radial force should dominate
both non--radial terms, leaving us with the strikingly simple,
one--dimensional equation of motion: \begin{eqnarray} {\ddot
r}_{1a}\ =\ - {1 \over \mu_a} u_1^\prime(r_{1a})\ .
\label{one-pair-dyn} \end{eqnarray}

In the event that both atoms of the solute are members of
mutual--nearest--neighbor pairs~\cite{pairings} with solvent atoms (the
situation described at the close of Sec. \ref{high-freq}, with the two
solvent atoms labeled $a$ and $b$), similar manipulations and arguments
tell us that the equation of motion should be generalized to the two
coupled differential equations \begin{eqnarray} {\ddot r}_{1a}\ &=&\ -
{1 \over \mu_a} u_1^\prime(r_{1a})\ - {1 \over M_{12}}
u_2^\prime(r_{2b}) \cos{\theta_{ab}} \cr\cr {\ddot r}_{2b}\ &=&\ - {1
\over \mu_b} u_2^\prime(r_{2b})\ - {1 \over M_{12}} u_1^\prime(r_{1a})
\cos{\theta_{ab}}\ , \label{two-pair-dyn} \end{eqnarray} with $\mu_b =
m_b M_{12} / (m_b + M_{12})$.  Nonetheless, consistent with our neglect
of rotational dynamics, we can regard the relative orientation of the
two pairs, $\cos{\theta_{ab}} \equiv ({\bf {\hat r}}_{1a} \cdot {\bf
{\hat r}}_{2b})$, as a constant of the motion, so the resulting
equations are still purely one--dimensional.  We might also predict
that, except for very rare liquid configurations, one of the two tagged
solvent atoms will be far more important than the other in determining
the force along the bond.  We therefore expect that we should never
have to do more than solve an equation of the form of Eq.
(\ref{one-pair-dyn}) for the most important solvent/solute--site pair.

\subsection{Putting the Pieces Together} \label{pair-theory-sec}

Because all we need to know for a given liquid configuration is the
location of the critical solvent closest to one end of the solute or
the other, our working formula for the vibrational relaxation
correlation function is simply, \begin{eqnarray} G_F(t)\ =\ {1 \over
4}\, \rho \, \int d{\bf r}\ g_{mnn}({\bf r};\ {\bf {\hat
r}_{12}})\ \left< u^{\prime\prime}[r(t)] u^{\prime\prime}[r(0)]
\cos^2{\theta(0)}\ {\dot r}(t) {\dot r}(0) \right>_{\bf r}\ ,
\label{g-from-struct} \end{eqnarray} for any homonuclear diatomic
dissolved in an atomic liquid.  Here $r$ is the distance between the
special solvent and the nearest solute atom, 
$g_{mnn}({\bf r};{\bf {\hat r}_{12}})$ is the probability density of
${\bf r}$, what we shall call the solute's
mutual--nearest--neighbor distribution function, and $\theta(0)$ is the
initial orientation of the solvent/solute--atom pair relative to the
bond direction, ${\bf {\hat r}_{12}}$.  The necessary dynamics launched from 
that configuration, $r(t)$, is then prescribed by the equation of
motion \begin{eqnarray} {\ddot r}(t)\ =\ - {1 \over \mu_{uv}} u'[r(t)]
\cr \cr r(0)\ =\ r ,\ {\dot r}(0) = v , \label{eqns-of-pair}
\end{eqnarray} with $\mu_{uv} = m M/(m + M)$ the solvent--solute
reduced mass ($m$ being the solvent atom's mass and $M$ being the
solute's total mass).

What liquid--structure effects there remain in the problem show up in
the function $g_{mnn}({\bf r};\ {\bf {\hat r}_{12}})$, \begin{eqnarray}
\rho g_{mnn}({\bf r};\ {\bf {\hat r}_{12}})\ =\ \left< \sum_{A = 1, 2}
{}^\prime \sum_{\hspace{.3cm} j > 2,\ j = mnn\ of\ site\ A} \delta[{\bf
r} - ({\bf r}_j - {\bf r}_A) ] \right> \label{define-mnn-dist}
\end{eqnarray} which is the {\it equilibrium} probability density for
finding a solvent atom that is both a mutual nearest neighbor and is
displaced by ${\bf r}$ from one end of the solute ($\rho$ being the
solvent number density).  In the event that both solute sites are
members of mnn pairs,~\cite{pairings} the primed summation selects the
site, $A$, whose partner is closest.  Consistent with this defintion,
by the brackets with subscript ${\bf r}$, we mean just the equilibrium
average over initial velocities $v$ in Eqs.  (\ref{g-from-struct}) and
(\ref{eqns-of-pair}), carried out at fixed ${\bf r}$.

The ultimate test of our approach, of course, will be its numerical
accuracy in reproducing the true high--frequency vibrational friction,
but even in advance of such tests it is worth looking at the internal
consistency of the theory.  Supposing for the moment that our basic
idea is correct --- that one or two critical solvents dominate the
instantaneous friction --- we still have to remember that the rest of
the liquid continues to evolve while the critical solvent is moving.
Is it even obvious that the particular solvent that makes up the
initial mnn pair with the solute remains an mnn pair throughout the
relevant time window?  As we shall see, for Xe liquid the
high--frequency part of the friction is largely determined by dynamics
occuring at times less than 300 fs.  A typical speed for the mnn
distance is $(k_B T/\mu_{uv})^{1/2}$, so at a temperature of 280 K, a
diatomic such as I$_2$ dissolved in Xe could have its mnn distance
change by about 0.5 $\AA$ during this time.  However, the mnn
distribution itself has a full width at half maximum that is typically
of this same order in dense Xe.~\cite{fwhm-i2}  It therefore seems
quite plausible that the identity of the mnn partner could be preserved
long enough for the theory to operate successfully.

\section{Instantaneous Normal Modes and Instantaneous--Pair Theory}
\label{sec3}

The ease with which we can perform calculations with Eqs.
(\ref{g-from-struct}) and (\ref{eqns-of-pair}) aside, we have not
really finished our task; there is still a conceptual gulf we need to
cross.  Our theory for high--frequency vibrational population
relaxation, does not, in itself, resolve the central conflict between
V--V and V--T pictures of the relaxation.  On the one hand, the
instantaneous--pair theory certainly seems to suggest that
high--frequency vibrations are converted into the liquid's
translational degrees of freedom.  Indeed, the whole motivation for
separating out the contributions of a single solvent is that the
special solvent highest on the repulsive wall of the intermolecular
potential completely overshadows any other solvent --- and the fact
that this sharply repulsive portion of intermolecular potentials is
largely what comes into play in the evaluation of the equations of
motion means that the dynamics is a case study in classical scattering
rather than an example of bound-state motion.  In short, the relevant
dynamics seems to resemble translation more than intermolecular
vibrations.  By the same token, however, it is difficult to reconcile
this stance with the noteable success of
resonant--vibrational--energy--transfer models for frequencies within
the liquid's band,~\cite{goodyear97,ladanyi98} especially given the
observations that there are binary features within the
instantaneous--normal--mode band.~\cite{larsen97,ladanyi98}  It would
not be impossible for the mechanism of vibrational relaxation to switch
as we go from one frequency regime to another, but such a pronounced
dichotomy would be especially problematic in the immediate vicinity of
the band edge.

One way to think about this dilemma is to ask what would happen if we
tried to construct an instantaneous--normal--mode treatment assuming
that all of the assumptions of the instantaneous--pair theory were, in
fact, correct.  That is, suppose we assume that the
vibrational--friction correlation really is given by the
instantaneous-pair approximation of Eq.  (\ref{g-from-struct}) and that
the dynamics really ought to be that prescribed by the matching
instantaneous--pair approximation of Eq.  (\ref{eqns-of-pair}).  Rather
than solving Eq. (\ref{eqns-of-pair}) exactly, an INM theory would make
an instantaneous harmonic estimate for the time evolution:  the
displacement from the initial pair separation would be written
as~\cite{buchner92} \begin{eqnarray}
r(t)\ -\ r(0)\ \equiv\ q(t)\ =\ \left( {f_0 \over \mu_{uv} \omega_0^2}
\right) \left( 1 - \cos{\omega_0 t} \right) + \left( {v_0 \over
\omega_0} \right) \sin{\omega_0 t}\ , \label{inm-dyn} \end{eqnarray}
where $\omega_0 \equiv \sqrt{u^{\prime\prime}[r(0)]/\mu_{uv}}$ is the
instantaneous frequency of the ``harmonic mode'' and $f_0 \equiv -
u^\prime[r(0)]$ is the instantaneous force.

However, besides just assuming that the dynamics was harmonic, the
traditional, linearized version of INM theory would go one, key, step
further.  The evolution of the coupling force, here $u^\prime(r)$,
would be linearized in the mode coordinate~\cite{stratt94}
\begin{eqnarray} u^{\prime}[r(t)]\ \approx\ u^{\prime}[r(0)] +
u^{\prime\prime}[r(0)]\ q(t)\ . \label{linforce} \end{eqnarray} This
step is what leads to the appealing form of both INM linearized
solvation theory and the INM theory for vibrational
friction.~\cite{goodyear96_1,goodyear96_2,goodyear97,larsen97,ladanyi98,kalbfleisch98,ladanyi96}
The appearance of what have been called the INM influence spectra stems
from this very step.  But it should be clear from our analysis of
instantaneous--pair theory that this step is strongly suspect under the
very conditions that lead to high-frequency relaxation.  If the
intermolecular potential is really sharply varying, then that is the
last place where one should attempt a low--order expansion.

We could raise the same kinds of doubts about the harmonic dynamics of
Eq. (\ref{inm-dyn}), but suppose, for the moment, we pursue just the
issue of nonlinearity in the coupling.  How would the nonlinearity
alone make itself felt?  We can present this question more generally if
we return to a general solute--centered correlation function introduced
in Sec. \ref{coup-dyn-sec}.  Within the framework of the
instantaneous--pair assumptions, the correlation function we need to
evaluate is of the form \begin{eqnarray} G_F(t)\ =\ \left<
w^\prime[r(t)]\ w^\prime[r(0)]\ {\dot r}(t)\ {\dot r}(0) \right>\ ,
\label{general-G} \end{eqnarray} where $w(r)$ is some spectroscopic
probe potential which may, but need not, be related to $u(r)$, the pair
potential governing the dynamics.

Were we to linearize $w(r)$ the same way that we did the force in Eq.
(\ref{linforce}), Eq. (\ref{inm-dyn}) would allow us to write the
correlation function as the cosine transform of an influence spectrum,
$\rho(\omega)$, a standard INM outcome. \begin{eqnarray}
&G&_F^{linear-INM}(t)\ =\ \left( k_B T \right) \int
d\omega\ \rho(\omega)\ \cos{\omega t}\ .  \label{g-from-infl-stuff}
\\ \cr &\rho&(\omega)\ =\ {1 \over \mu_{uv}} \left< \left( w_0^\prime
\right)^2 \delta ( \omega - \omega_0 ) \right>\ .  \label{infl-spect}
\end{eqnarray}  Here $\mu_{uv}$ is the appropriate reduced mass,
$w_0^\prime \equiv w^\prime[r(0)]$, $\omega_0 \equiv
\sqrt{u^{\prime\prime}[r(0)]/\mu_{uv}}$, and the brackets denote the
(fully anharmonic) equilibrium average over the initial coordinate
$r(0)$.  As we might have expected, however, this influence spectrum is
too limited; it spans a range of frequencies no larger than the INM
band (the density of states) itself \begin{eqnarray}
D(\omega)\ =\ \left< \delta (\omega - \omega_0) \right>\ . \label{DOS}
\end{eqnarray}  We want to point out, though, that the
instantaneous--pair formulation of the problem makes it simple enough
to include the full nonlinearity of the coupling, even while retaining
the harmonicity of our dynamics.  If we carry out the equivalent of
Eq.  (\ref{linforce}) to all orders \begin{eqnarray}
w^\prime[r(t)]\ =\ \sum_{n=0}^\infty {1 \over n!}
w_0^{(n+1)}\ q^n(t)\ , \end{eqnarray} with $w_0^{(n)}$ the n--th
derivative of $w(r)$ evaluated at the instantaneous position $r(0)$, we
can still substitute Eq. (\ref{inm-dyn}) into Eq. (\ref{general-G}),
providing us with an expression for the necessary correlation function
in terms of an infinite sum \begin{eqnarray}
&G&_F^{INM}(t)\ =\ \sum_{n=1}^\infty G_F^{(n)}(t)\ , \\ \cr
&G&_F^{(n)}(t)\ =\ {1 \over (n-1)!}\ \left<
w_0^{(n)}\ w_0^{(1)}\ q^{n-1}(t)\ {\dot q}(t)\ {\dot q}(0) \right>\ ,
\label{g-super-n} \end{eqnarray} though the bracket now represents the
average over the initial velocity as well as the anharmonic average
over the initial coordinate.  But, as we show in the Appendix, this
velocity average can be performed analytically, yielding the result
that the n--th term is \begin{eqnarray} G_F^{(n)}(t)\ &=&\ {k_B T \over
\mu_{uv}}\ \left< w_0^{(n)} w_0^{(1)}\ V_n[t;\ r(0)] \right>\ , \cr\cr
V_n[t;\ r(0)] &=& \left[ {-i Y(t) \over \alpha_0} \right]^{n-1} \left\{
( n \cos{\omega_0 t}) H_{n-1}[i X(t)]\ +\ (n-1) [2 i X(t)] H_{n-2}[i
X(t)] \right\}\ , \cr\cr X(t)\ &=&\ \left( {\omega_0 t_0 \over 2}
\right)\ \tan{{\omega_0 t \over 2}}\ ,\ Y(t)\ =\ { \sin{\omega_0 t}
\over \omega_0 t_0}\ .  \label{hermite-sum} \end{eqnarray}  In these
formulas $H_n(x)$ is the n--th order Hermite polynomial~\cite{abram72}
and we have defined the chararacteristic length scale and time scale
for the dynamics by the instantaneous--position--dependent quantities
\begin{eqnarray} \alpha_0\ =\ {\mu_{uv} \omega_0^2 \over f_0}\ ,\quad
t_0\ =\ \left( {2 \mu_{uv} \over k_B T \alpha_0^2 } \right)^{1/2}\ ,
\label{def-params} \end{eqnarray} respectively.

The first term in this series is, of course, the standard linearized
INM result, \begin{eqnarray} G_F^{(1)}(t)\ =\ G_F^{linear-INM}(t)\ ,
\end{eqnarray} involving a configurational average of (something
proportional to) $\cos{\omega_0 t}$, but each higher order term clearly
brings in higher powers of the trigonometric functions.  In fact, since
$H_n(x) = (2 x)^n$ plus lower powers in $x$,~\cite{abram72} it is easy
to show that each $V_n$ integral is a polynomial form in $\cos{\omega_0
t}$ and $\sin{\omega_0 t}$, and that the leading (highest order) term
is proportional to $\cos^n{\omega_0 t} = \cos{ n \omega_0 t}\ +$ lower
order harmonics of $\omega_0$.  In other words, each successive term
brings in successively higher--order overtones of the fundamental INM
band of the liquid.  The instantaneous pair theory with INM dynamics is
thus a genuine multi--phonon theory of high-frequency vibrational
relaxation in liquids --- as long as the proper respect is paid to the
nonlinearity of the coupling.~\cite{multi-phon,liq-analog}

We should hasten to point out that a literal term by term evaluation of
this sum is not necessarily going to be of any great computational
help.  For one thing, it is not straightforward to write the n--th
order equivalent of the (one--phonon) influence spectrum, Eq.
(\ref{infl-spect}).  Since $V_n$ not only has terms of the form $\cos{
n \omega_0 t}$, but can include $\cos{ (n-1) \omega_0 t},\ \cos{ (n-2)
\omega_0 t}, \ldots,$ and $\cos{ \omega_0 t}$ terms as well, the final
$n$--phonon spectrum could have contributions from $V_n$, $V_{n+1}$,
and all of the higher--order integrals.  A more fundamental way of
making this same point is to remember that there is no real reason to
expect that any low--order truncation of the multi--phonon series will
adequately describe the relaxation process.  A more realistic approach
to the problem might require us to sum the series to infinite order, at
least in some approximate fashion.

It is straightforward enough to evaluate the combination of Eqs.
(\ref{inm-dyn}) and (\ref{general-G}) numerically, without making any
approximations, but it is revealing to examine a special case for which
we can perform the necessary multiphonon sum analytically.  If the
spectroscopic probe potential is an exponential repulsion (a fairly
common assumption in theories of vibrational
relaxation)~\cite{egorov95_2,miller94,nitzan75} \begin{eqnarray}
w[r(t)]\ =\ w[r(0)]\ e^{-\alpha \left[ r(t) - r(0) \right] }\ ,
\label{exp-probe} \end{eqnarray} with $\alpha$ a constant, then all of
the derivatives in Eq. (\ref{hermite-sum}) can be expressed in terms of
$\alpha$ and $w_0^\prime \equiv w_0^{(1)}$. \begin{eqnarray}
&G&_F(t)\ =\ \left( {k_B T \over \mu_{uv} } \right) \left<\ \left(
w_0^\prime \right)^2 V[t;\ r(0)]\ \right>\ , \cr\cr
&V&[t;\ r(0)]\ =\ \sum_{n=1}^\infty V_n[t;\ r(0)]
{\left(-\alpha\right)^{n-1} \over (n-1)!}\ . \label{exp-sum}
\end{eqnarray}  As detailed in the Appendix, the sum can now be
evaluated in closed form in terms of the ratio $\gamma_0 \equiv
\alpha/\alpha_0$. \begin{eqnarray}
V[t;\ r(0)]\ &=&\ S[t;\ r(0)]\ e^{R[t ;\ r(0)]}\ , \cr\cr
R[t;\ r(0)]\ &=&\ \gamma_0 \left( \cos{\omega_0 t} - 1
\right)\ +\ \gamma_0^2 \left( { \sin{\omega_0 t} \over \omega_0 t_0}
\right)^2\ , \cr\cr S[t;\ r(0)]\ &=&\ \cos{\omega_0 t}\ -\ \gamma_0
\sin^2{\omega_0 t}\ +\ 2 \gamma_0^2 \cos{\omega_0 t} \left(
{\sin{\omega_0 t} \over \omega_0 t_0 } \right)^2\ . \label{summed-form}
\end{eqnarray}  In fact, for vibrational relaxation the situation
becomes even simpler.~\cite{anharm-stuff}  If the interatomic potential
itself is exponential \begin{eqnarray} u[r(t)]\ =\ u[r(0)]\ e^{-\alpha
[r(t) - r(0)]}\ ,\label{exp-pot} \end{eqnarray} then $w(r) =
u^\prime(r)$ is proportional to the same exponential.  The ratio
$\gamma_0 \equiv \alpha/\alpha_0$ is thus identically equal to 1 and
the charateristic time reduces to the constant \begin{eqnarray}
t_0\ =\ \left( { 2 \mu_{uv} \over k_B T \alpha^2 } \right)^{1/2}\ .
\label{t0-def} \end{eqnarray}

Equations (\ref{exp-sum}), (\ref{summed-form}), and (\ref{t0-def})
constitute our basic non--linear INM theory for high--frequency
vibrational relaxation.  We shall see presently how well it fares
compared to nonlinear--coupling plus anharmonic--dynamics treatment of
the full instantaneous--pair theory.

\section{Numerical Results}

\subsection{Numerical Methods and Models}

In this section we numerically evaluate the vibrational friction felt
by a rigid homonuclear diatomic molecule dissolved in an atomic
liquid.  The atoms of the solvent are taken to interact with each other
and with the two atoms of the solute through Lennard--Jones pair
potentials \begin{eqnarray} u(r)\ =\ 4 \epsilon \left[ \left({ \sigma
\over r} \right)^{12} - \left({ \sigma \over r} \right)^6 \right]  ,
\nonumber \end{eqnarray} though the solvent atom--solvent atom ($vv$)
and solute atom--solvent atom ($uv$) Lennard--Jones parameters can
differ, as can the masses of the solute and solvent atoms ($m_u$ and
$m_v$, respectively).  While the effect of the solvent on the
intramolecular vibration is the whole point of these calculations, for
the purposes of computing the friction, the solute atoms can be
regarded as being separated by a fixed distance $r_{eq}$.

For purposes of definiteness, we regard our solvent as fluid Xe
($\sigma_{vv} = 4.1\ \AA$, $\epsilon_{vv} = 222\ K$, $m_v = 131.3$
amu).~\cite{lj-xe-ref}  The two specific solution models we consider
are the standard Tuckerman and Berne model,~\cite{berne90,tuckerman93}
in which the two solute atoms are regarded as identical to the solvent
atoms, \begin{eqnarray}
\sigma_{uv}\ =\ \sigma_{vv}\ ,\ \epsilon_{uv}\ =\ \epsilon_{vv}\ ,\ m_u\ =\
m_v\ ,\ r_{eq}\ =\ 1.25 \sigma_{vv}\ , \nonumber \end{eqnarray} and the
model used by Brown, Harris, and Tully,~\cite{brown88} \begin{eqnarray}
\sigma_{uv}\ =\ 3.94\ \AA,\ \epsilon_{uv}\ =\ 324\ K,\ m_u\ =\ 126.9\
\mbox{amu},\ r_{eq} =  2.67\ \AA\ . \nonumber \end{eqnarray}  This
latter model was explored largely in response to experimental studies
by Paige, Russell and Harris of the relaxation of highly
vibrationally--excited $I_2$ dissolved in Xe.~\cite{paige90}

Our theoretical predictions were constructed so as to rely solely on
the equilibrium properties of the solutions being studied.  We  were
therefore able to obtain the information we needed by using standard
canonical--ensemble Monte Carlo simulations~\cite{allen87} with 106
solvent atoms and an immobile diatomic solute.  All of these
simulations employed periodic boundary conditions and had their
maximum--move distances adjusted to achieve a 50\% acceptance rate for
the Metropolis attempted moves.  Ensemble averages were then computed
by sampling fluid configurations every $N_t$ such attempts, with $N_t$
chosen so as to ensure minimal configuration--configuration correlation
in the force along the diatomic's bond.  The actual values of $N_t$
ranged from a low of 2000 for dense gases to as high as 5000 for
high--density fluids.  When exact dynamical results were needed to give
us something with which we could compare, we carried out full molecular
dynamics simulations on the same rigid--diatomic--plus--106--solvent
system, now allowing the diatomic to translate and rotate.
Specifically, classical trajectories of length
$\tau_{LJ}\ [\tau_{LJ}\ =\ (m_{vv}
\sigma_{vv}^2/\epsilon_{vv})^{1/2}\ =\ 3.47 ps]$ were computed starting
from each of $10^5$ (uncorrelated) Monte--Carlo generated
configurations using the Rattle version of the velocity--Verlet
integrator~\cite{verlet} with a time step of 0.001 $\tau_{LJ}$.  The
initial velocities for these trajectories were selected from the
Maxwell--Boltzmann distribution.~\cite{init-vel}

Results for the pair theories were obtained in much the same way as the
exact molecular dynamics, but the requisite trajectories involved only
the motion of one--dimensional solute--solvent distances $r(t)$.  The
outcome for the full instantaneous--pair theory could therefore
calculated by using a simple velocity--Verlet integration starting from
the Monte--Carlo generated liquid--state
configurations.~\cite{verlet,one-dim}  Averaging over the initial
velocities was then accomplished by using 40 velocities equally spaced
between $\pm (40k_BT/\mu_{uv})^{1/2}$, appropriately weighting by the
Maxwell--Boltzmann distribution.  The computation with the alternative
pair theory, the corresponding nonlinear--INM approach, was even
easier, since it only needed the analytical dynamics of Eq.
(\ref{inm-dyn}), though it too required numerical averaging over the
initial velocities.

All cosine transforms were performed using standard FFT
methods.~\cite{num-rec}

\subsection{Basic Results}

\begin{figure}

\epsfig{file=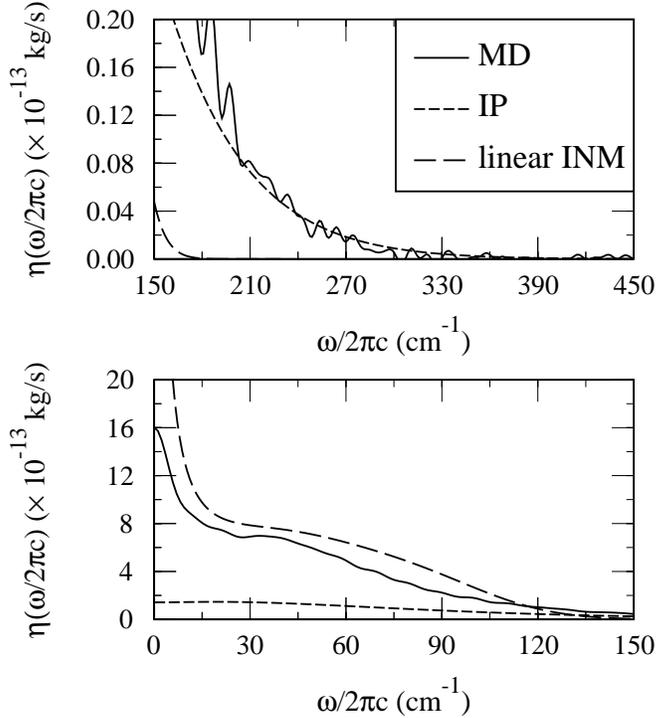,height=12cm}

\caption[caption]{The vibrational friction felt by a
vibrating diatomic molecule dissolved in a supercritical atomic
liquid.  Shown here are a number of different calculations for the real
part of the frequency--domain vibrational friction of the
Tuckerman--Berne model system (under the standard model conditions of
reduced density $\rho \sigma_{vv}^3 = 1.05$ and reduced temperature
$k_B T / \epsilon_{vv} = 2.5$) in units appropriate to a Xe solvent.
The lower panel is the friction relevant to diatomic whose vibrational
frequency falls within the expected bandwidth of the solvent; the upper
panel portrays the friction for frequencies above the band edge.  (Note
the rather different scales used in the two panels.)  In both panels,
the rough solid line results from an exact molecular dynamics
calculation of the bond--force autocorrelation function, the long
dashed line (visible only in the lower left hand corner in the upper
panel) shows the analogous predictions from linearized
instantaneous--normal--mode theory, and the short dashed line gives the
analogous predictions from full instantaneous--pair theory.}
\label{LJ-fric-fig} \end{figure}

We begin our exploration by looking at the vibrational friction felt in
the Tuckerman and Berne model of a diatomic solute dissolved in a high
density supercritical fluid.  As one can see from the bottom panel of
Fig.  \ref{LJ-fric-fig}, our earlier linearized--INM theory (the
long--dashed line) does a reasonably credible job of reproducing the
shape of the exact--molecular--dynamics--derived friction (the solid
line) within the band of the liquid.  However once we venture beyond
the band edge (upper panel) the discrepancy between the two becomes
painfully obvious.  By contrast, the full instantaneous pair theory
(the short--dashed line) provides a virtually quantitative match to the
exact results at these higher frequencies.  Apparently combining a
nonlinear treatment of the solvent coupling with an anharmonic approach
to the solvent dynamics amply suffices to capture the missing
high--frequency response of the solvent --- and it does so despite our
omission of all but an infinitesmal fraction of the solvent dynamics.
The relatively poor performance of the same instantaneous--pair theory
within the solvent band (lower panel) continues to remind us, at the
same time, that an accurate treatment of the low--frequency regime
requires a proper inclusion of the more collective aspects of the
solvent's motion.

These same points are made more explicitly by using Eq.
(\ref{l-t-form}) to compute the actual vibrational energy relaxation
rates from the friction.  As we can see from Table \ref{rel-times}, the
instantaneous--pair theory precisely mimics the
two--orders--of--magnitude drop in relaxation rate that transpires once
we go from vibrations in the middle of the liquid's band to those 100
$cm^{-1}$ outside the band.  Yet, within the band, the rate predicted
by this theory is consistently too small, just what we might have
surmised would happen if we were leaving out the dynamical channels for
relaxation provided by many--body motions.  Linearized INM theory, on
the other hand, behaves in a perfectly complementary fashion.  It gives
sensible relaxation rates within the band, but can do no better than
trace out the precipitous decay of the band edge once we start to leave
the band.

\begin{figure} 

\epsfig{file=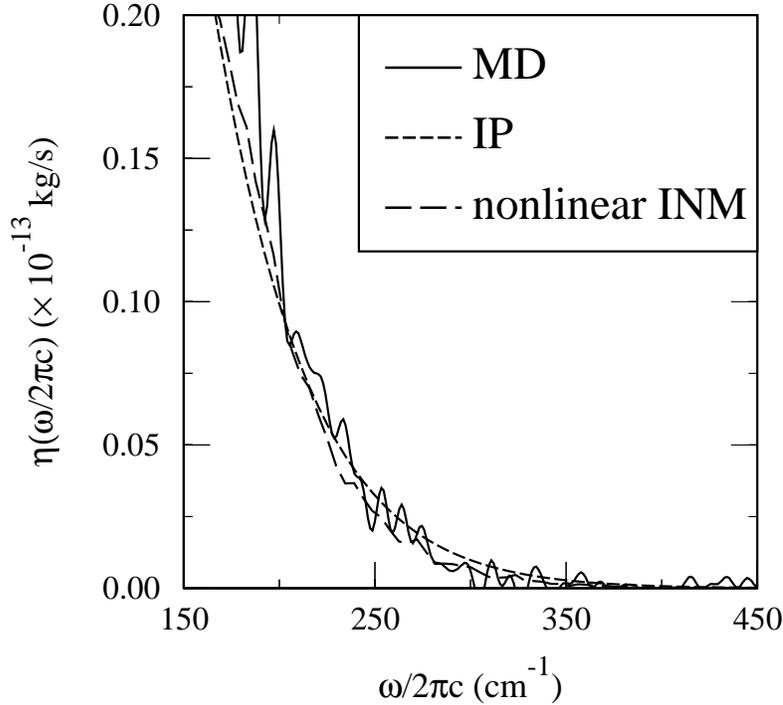}

\caption[caption]{The vibrational friction felt by a
high-frequency vibrating diatomic molecule dissolved in a supercritical
atomic liquid.  As in the upper panel of Fig. \ref{LJ-fric-fig}, the
calculations are performed for the standard Tuckerman--Berne model
system (under the same thermodynamic conditions) with Xe as the
solvent, and the solid and short-dashed lines show the
full--molecular--dynamics and the full instantaneous--pair theory
results, respectively.  Here, however, the long--dashed line represents
the outcome of the nonlinearly--coupled instantaneous--normal--mode
version of the pair theory.  Both pair theories interpret the
relaxation as a few--body event in which only the initial conditions
are prescribed by the fluid.  They differ in that the full
instantaneous--pair--theory uses the anharmonic dynamics of the key
atoms to drive the evolution of the nonlinear solute--solvent coupling,
whereas the INM equivalent assumes harmonic dynamics drives this
coupling.  In contrast to both of these treatments, the linear INM
theory of Fig. \ref{LJ-fric-fig} limits itself to linear coupling.}
\label{inm-fric-fig} \end{figure}

Our next question, clearly, is whether the success of the full pair
theory means that a harmonic approach to the solvent dynamics suddenly
becomes inappropriate at high frequencies.  Looking again at the
high--frequency friction, Fig. \ref{inm-fric-fig}, we find that if
retain the exact nonlinearity in Eq. (\ref{g-from-struct}), but use
instantaneous--normal--modes to prescribe the
mutual--nearest--neighbor--pair dynamics, Eq. (\ref{inm-dyn}), the
resulting friction spectrum is still virtually quantitative in its
agreement with the exact molecular dynamics.  As we might have
anticipated from the multiphonon discussion in the last Section, having
nonlinear coupling allows even harmonic dynamics to span a realistic
frequency range.  What we might not have predicted, however, was that
the coupling was so much more important than the anharmonicity that
incorporating the nonlinearity of the coupling was all one needed to do
to recover accurate vibrational relaxation rates.~\cite{bader96}

The level of similarity between these two pair theories is actually
even greater than it might appear on the scale of Fig.
\ref{inm-fric-fig}.  One of the more dramatic findings of the studies
of vibrational relaxation in the solid state is that once the amount of
energy transferred gets large enough, relaxation rates tend to obey an
exponential gap law.~\cite{egorov95_1,egorov95_2,nitzan74,nitzan75}
That is, the rates depend on the frequency of the solute vibration
$\omega$ roughly as \begin{eqnarray} T_1^{-1}\ \sim\ \exp{- \left(
{\omega \over \omega_{char} } \right)}\ ,\ \ (\omega \gg\ \omega_{char}
)\ ,\label{exp-gap} \end{eqnarray} where $\omega_{char}$ is some
characteristic phonon frequency of the solid.  There is a long history
to these kinds of exponential relations for a variety of different
kinds of energy transfer processes,~\cite{exp-gap} but the particular
application to solids is well known to follow directly from a
multiphonon picture of the relaxation:  steepest--descent evaluation of
the relaxation rate predicts the rate to be exponential in the number
of phonons excited.

\begin{figure}

\epsfig{file=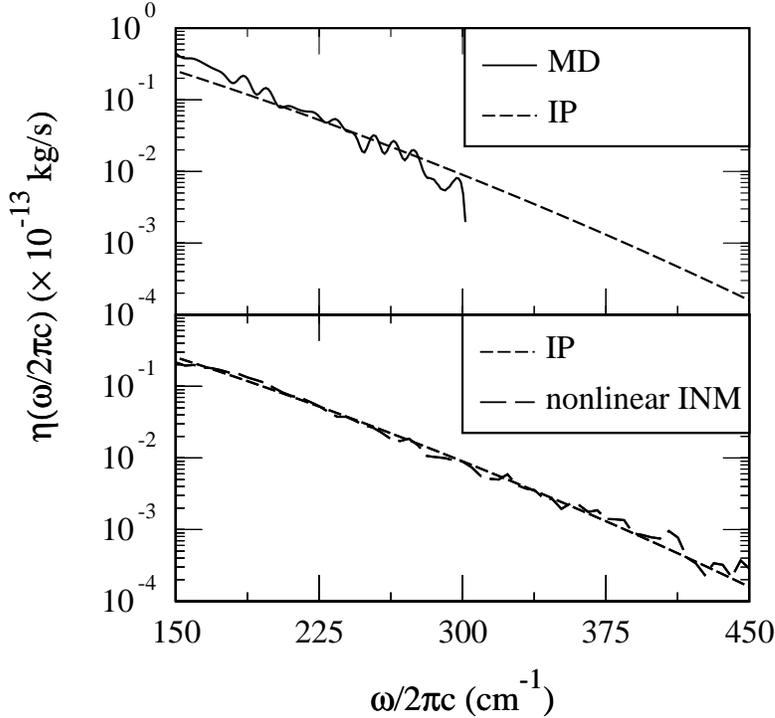}

\caption[caption]{The vibrational friction felt by a
high--frequency vibrating diatomic molecule dissolved in supercritical
atomic Xe.  In the upper panel, the instantaneous--pair (short dashes)
and molecular dynamics (solid line) results from Fig.
\ref{inm-fric-fig} are replotted here on a semilog scale to emphasize
the near exponential--gap--law behavior.  The lower panel compares the
instantaneous--pair theory of the upper panel to the analytical
results, Eqs. (\ref{exp-sum}), (\ref{summed-form}), and (\ref{t0-def}),
where to generate $t_0$ we have used the value of $\alpha_0$ found by
averaging over initial mnn distances.  For the supercritical fluid,
this average $\alpha_0$ is $17.2\ \sigma_{Xe}^{-1}\ =\ 4.2\ \AA^{-1}$,
which yields a characteristic timescale,
$t_0\ =\ 0.042\ \tau_{LJ}\ =\ 146 fs$.  Numerical noise clearly makes
it difficult to follow the exact molecular dynamics results over more
than a limited frequency interval.  It is only by going beyond this
interval, however, that the small deviations from strictly exponential
behavior begin to manifest themselves in the form of a slight downward
curvature to the predictions of the pair theories.}
\label{log-fric-fig} \end{figure}

What happens in liquids?~\cite{egorov96,nitzan97}  A logarithmic plot
of the relaxation rate, Fig. \ref{log-fric-fig}, shows that the decline
in the molecular--dynamics--computed rate with solute frequency is
indeed close to exponential, a trend reproduced exceptionally well by
the instantaneous pair theory.  In fact, given the numerical
difficulties inherent in extracting the high--frequency components of a
molecular--dynamics trajectory,~\cite{nitzan97} we might venture to
suggest that the pair theory results are actually the more reliable of
the two at high frequencies.~\cite{one-dim,pair-stable}  Interestingly,
the exponential line that the two curves seem to follow is
\begin{eqnarray} T_1^{-1}\ \sim\ \exp{ -\left( \omega t_0 \right) }
\end{eqnarray} with $t_0$ the very time scale prescribed by nonlinear
INM theory, Eq.  (\ref{def-params}).  This statement actually requires
a little qualification.  Our formula for $t_0$ expresses it as a
function of $\alpha_0$, which, itself is a function of the initial mnn
distance $r(0)$.  However we can evaluate an average $\alpha_0$ {\it a
priori} by using the $\alpha_0$ values of Eq.  (\ref{def-params}) for
the solute--solvent Lennard--Jones potential averaged over the
equilibrium distribution of initial mnn distances, Eq.
(\ref{define-mnn-dist}).  The resulting $\alpha_0$ is what is used to
generate the $t_0$ for the long--dashed line in the
figure.~\cite{alpha-ref}

Of course, coming back to the solid--state analogy, we could regard an
exponential gap law coming out of nonlinear INM predictions as
completely consistent with our multiphonon--like interpretation of the
INM theory.  In fact, it is not hard to show that our simple analytical
results of Eqs. (\ref{exp-sum}), (\ref{summed-form}), and
(\ref{t0-def}) do predict the same exponential decay as the full
instantaneous pair theory (Fig.  \ref{log-fric-fig}, bottom panel).
Apparently, even with this intrinsically vibrational picture of the
solvent dynamics embedded in the calculation, the agreement between the
pair theories with harmonic and anharmonic dynamics is robust enough to
span a number of decades of solvent response.

\begin{figure}

\epsfig{file=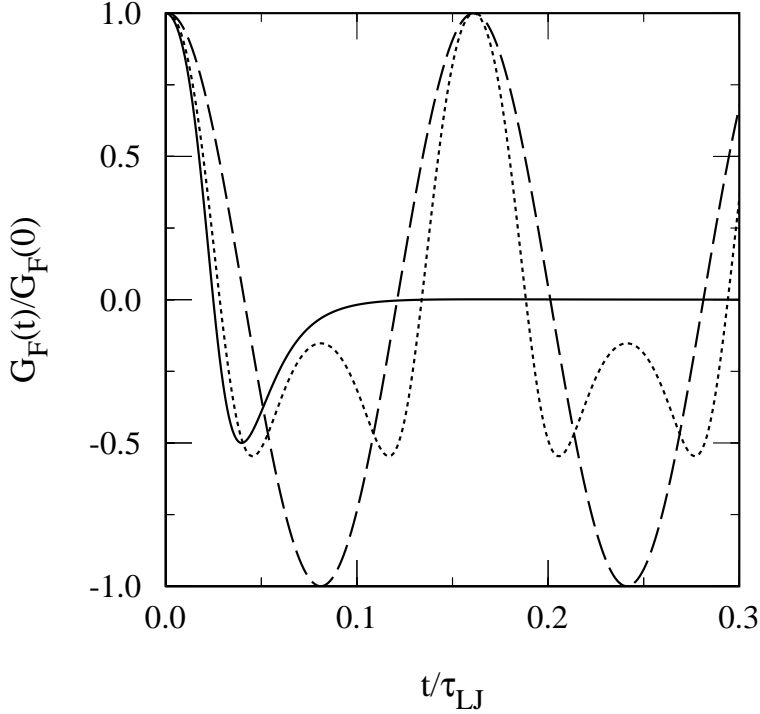}

\caption[caption]{The bond--force--velocity
autocorrelation function for the standard Tuckerman--Berne model system
evaluated at a single initial liquid configuration for three different
pair theories.  Shown here is a normalized version of the instantaneous
correlation function appearing in Eq. (\ref{g-from-struct}), $<
u^{\prime\prime}[r(t)]\ u^{\prime\prime}[r(0)] \cos ^2 \theta_0\ {\dot
r}(t) {\dot r}(0) >_{{\bf r_0}}$, with the initial displacement between
the key solute and solvent atoms held fixed at $r_0 = | {\bf r_0} | =
0.95 \sigma_{uv}$ and $\theta_0 = 0$.  As in Figs.
\ref{LJ-fric-fig}--\ref{log-fric-fig}, the reduced temperature $k_B T/
\epsilon_{vv} = 2.5$.  The solid and dotted lines both make use of the
manifestly nonlinear solute--solvent Lennard--Jones potential $u(r)$ in
evaluating $u^{\prime\prime}(r)$, but they take the time evolution to
be given by the anharmonic dynamics of Eq.  (\ref{eqns-of-pair}) and
the harmonic dynamics of Eq.  (\ref{inm-dyn}) respectively.  The long
dashed curve results from the corresponding linear INM theory:  that
is, from employing both the harmonic dynamics and the linear coupling
assumption $u^{\prime\prime} [r(t)]\ \approx\ u^{\prime\prime}[r(0)]$.
For Xe, a time interval of $0.1\ \tau_{LJ}$ corresponds to about 350
fs.} \label{gf-time-fig} \end{figure}

It is revealing to think in some detail about this remarkable agreement
between the full instantaneous pair theory and its
nonlinear--coupling/harmonic--dynamics analogue.  A closer inspection
of the two theories in the time domain, Fig. \ref{gf-time-fig},
emphasizes the apparently fundamental differences between the two views
of the dynamics.  If we restrict ourselves to a single initial liquid
configuration, as the figure does, we see that the harmonic curve
reflects the oscillatory character of bound--state motion, whereas the
lack of periodicity in the full (anharmonic) result is consistent with
what the pair dynamics really is --- a scattering process.  So why
should such disparate views lead to the same results?  The answer is
suggested by the third curve in Fig. 4, which also assumes harmonic
dynamics but truncates the coupling at linear order, precisely the sort
of theory we would configurationally average to arrive at Eqs.
(\ref{g-from-infl-stuff}) and (\ref{infl-spect}).  It is evident that
neither of the two harmonic curves provides all that faithful a
rendering of the anharmonic dynamics, but what stands out is that the
nonlinear coupling allows for a much more accurate reconstruction of
the correlation function at short times, reproducing not only the
initial decay of the exact correlation function, but also its one
oscillation.~\cite{inertial}

It certainly makes sense that the highest frequency behavior of the
friction depends critically on the shortest time dynamics, but we can
say a little more than that.  Obviously, to get the correct
high--frequency Fourier transform of $G_F(t)$, it is important that we
know its behavior everywhere it changes rapidly --- which includes both
the initial fall--off and the subsequent rebound we just referred to.
Equally important, though, is the fact that when these
single--configuration correlation functions are averaged, the spurious
oscillatory behavior of the harmonic picture that we see at longer
times, which seems to portray solvent molecules as colliding with the
solute over and over again, dephases away to nothing.  Thus, at least
within the time window provided by the rapid decay of the mnn
bond--force--velocity correlation, vibrational motion can indeed manage
to look identical to single scattering events.  To put it another way,
for our own limited purposes of computing vibrational population
relaxation rates, we can evidently continue to regard the essential
liquid dynamics as intermolecular vibrations, both at high and at low
frequencies.

\subsection{Thermodynamic State Dependence and the Application to $I_2$
in Xe}

We close our discussion with a look at a more experimentally oriented
calculation, that for the vibrational population relaxation of the
first vibrationally--excited state of $I_2$ dissolved in fluid Xe at 280
K.~\cite{egorov96,paige90}

\begin{figure}

\epsfig{file=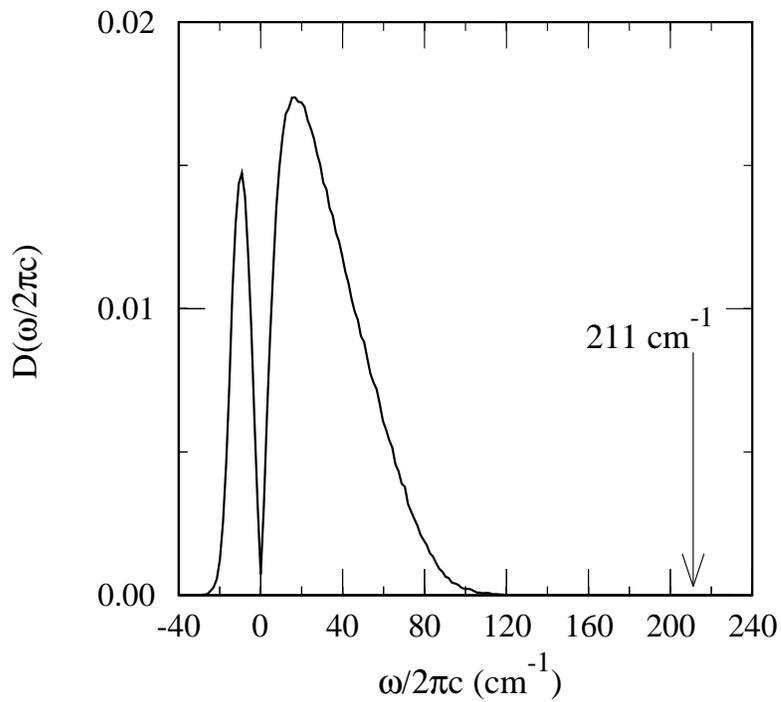}

\caption[caption]{The instantaneous--normal--mode
density of states for liquid Xe at a density $\rho = 3.0 g/ml$ and
temperature $T = 280 K$ (computed by taking Xe to be a Lennard--Jones
liquid at reduced density and temperature $\rho \sigma^3 = 0.95$ and
$k_B T/\epsilon = 1.26$).  Shown by the arrow, for comparison, is the
vibrational frequency of I$_2$ in its ground vibronic state.}
\label{xe-band} \end{figure}

To begin with, we should note that from xenon's perspective, the $I_2$
vibrational frequency of $(\omega / 2 \pi c) = 211\ cm^{-1}$ actually
qualifies as a rather high frequency.  As is clear from Fig.
\ref{xe-band}, this frequency lies well outside the INM band of Xe
liquid (and even further outside the band of lower density Xe).  It
therefore confronts us with precisely the issues we have been
discussing.  Another interesting feature of this example, though, is
that at this temperature, Xe can be taken from dense liquid, through
the liquid--gas coexistence region, into a rarified gas.  We can
therefore examine a range of dynamical possibilities within this one
example.  The only additional step we need to take is to provide some
way to compute the relaxation rates within the coexistence region using
data taken only from single--phase, constant temperature and volume,
simulations.

To meet this last requirement, we assume that the rate at which a
solute relaxes vibrationally, $T^{-1}$, is far faster than the rate at
which it moves between liquid and gaseous regions.  Under these
conditions, the experimentally measured relaxation rate should be
simply the weighted sum \begin{eqnarray} T^{-1}\ =\ x_{liq} \left(
T^{-1} \right)_{liq}\ +\ x_{gas} \left( T^{-1} \right)_{gas}\ ,
\label{coex-form} \end{eqnarray} where $(T^{-1})_{liq}$ and
$(T^{-1})_{gas}$ are the relaxation rates calculated along the
coexistence line in the liquid and gaseous phases, respectively, and
$x_{liq}$ and $x_{gas}$ are the respective mole fractions of the two
phases at the given temperature and total number density $\rho$.  These
mole fractions, in turn, are derived from the lever rule
\begin{eqnarray} x_{liq}\ =\ 1\ -\ x_{gas}\ =\ { \rho\ -\ \rho_{gas}
\over \rho_{liq}\ -\ \rho_{gas} }\ , \label{lever-rule}\end{eqnarray}
and the values of $\rho_{liq}$ and $\rho_{gas}$, the individual
densities of the liquid and gas phases.

\begin{figure}

\epsfig{file=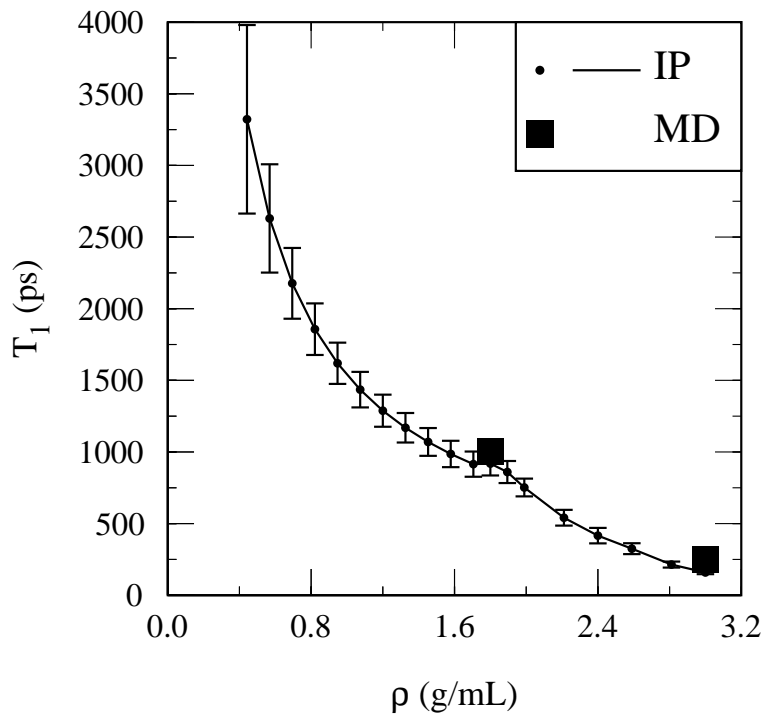}

\caption[caption]{Predicted vibrational population
relaxation times, $T_1$, as a function of liquid density, $\rho$, for
I$_2$ dissolved in liquid Xe at 280 K.  Results from the full
instantaneous--pair--theory are indicated by dots with a line
connecting them drawn to guide the eye.  These values were obtained by
averaging the instantaneous results over 100,000 liquid configurations
for densities less than 2.0 g/mL, 50,000 configurations for $\rho$ =
2.2 g/mL, and 20,000 configurations for all higher densities; 2
standard--deviation error bars are shown.  The point indicated by
squares are from the full molecular dynamics simulation results of
Brown, Harris, and Tully (Ref.  \onlinecite{brown88}) as reported by
Harris, Smith, and Russell (Ref. \onlinecite{harris90}).}
\label{I2-relax-fig} \end{figure}

Accordingly, if we use the full instantaneous pair theory to evaluate
the relaxation rates in the liquid and the gas, and we allow Eqs.
(\ref{coex-form}) and (\ref{lever-rule}) to interpolate between the
liquid (1.7 g/mL) and gaseous (0.44 g/mL) coexistence
densities,~\cite{coex-ref} we find that we can follow the vibrational
lifetime of $I_2$ over a wide range in solvent density (Fig.
\ref{I2-relax-fig}).  The general behavior shown here, that vibrational
lifetimes decrease monotonically with density, is hardly surprising,
but a number of features of this graph are worthy of some attention.
For one thing, we note that the simple pair theory predicts lifetimes
that are essentially identical to those computed by the elaborate
molecular dynamics simulations of Brown, Harris, and Tully, at least at
the two densities that those authors were able to
study.~\cite{brown88}  However, the computational simplicity of the
pair theory makes it straighforward for us to trace the overall trends
with thermodynamic state as well.  The basic increase in relaxation
rate that we see with density (a 16--fold increase over the same
interval where the density increases by roughly a factor of 6) is
something that one would have expected based in
independent--binary--collision (IBC)
models.~\cite{chesnoy86,paige90,dardi88}  Those models, though, would
attribute the increase to a hypothetical growth in the ``collision
rate''.  In the instantaneous--pair theory, by contrast, the solvent
density influences only the equilibrium probability of having a given
mutual--nearest--neighbor pair distance; higher densities translating
into smaller average distances.  The subsequent dynamics --- and
therefore the relaxation rate for that mnn distance --- is controlled
solely by the temperature, which sets the initial velocity distribution
of the solute--solvent pair, but which we hold fixed throughout the
calculation here.

The similar experimental consequences of the two different theoretical
models make it difficult to distinguish them just by measuring
relaxation rates versus density.  We emphasize, however, the IBC
model's reliance on an ill--defined collision--radius
parameter,~\cite{chesnoy86,paige90,dardi88} something the more
microscopically defined instantaneous--pair theory has no need for.  In
all fairness, we should also point out that both models lead to results
that are extraordinarily sensitive to the precise shape of the
repulsive wall of the solvent--solute potential.  Indeed, the
potentials employed here (and by Brown et al)~\cite{brown88} predict
relaxation rates differing by about a factor of 5 from the experimental
estimates.~\cite{egorov96}  That the likely origin of this discrepancy
does lie in the intermolecular potentials is supported by our finding
that small changes in the repulsive portion of the potential can shift
our relaxation times by as much as an order of magnitude.  This
sensitivity implies that we could tune our results to be more in accord
with the experimental data,~\cite{egorov96} but since such curve
fitting is not particularly germane to the issues at hand, we will
leave the matter here.

\section{Concluding Remarks}

The overriding goal of this paper was not really to find out how
quickly dissolved molecules can get rid of large quantities of
vibrational energy, but to learn how we should visualize such processes
in molecular detail.  To that end we deliberately phrased the issues in
terms of somewhat oversimplified caricatures.  Genuine molecular motion
in liquids is neither the simple vibrations of crystals nor the
occasionally interrupted ballistic motion found in gases.  What we
envisioned, though, was that keeping these two limiting cases in mind
would serve to frame the possibilities.  Even if the literal truth
ended up being somewhere in the middle, we presumed that we would still
be able to say whether molecules {\it tended} to lose vibrational
energy by finding a matching vibration--like motion somewhere in the
solvent (a more--or--less V--V event) or whether it was better to
regard relaxing solutes as simply adding generic translational kinetic
energy to the solvent (mostly a V--T process).

Much to our surprise, the real situation turned out to be far more
narrowly defined than we anticipated.  In effect, we found that the
structural disorder of liquids conspires with the short--ranged,
rapidly--varying character of repulsive intermolecular forces to
prescribe a remarkably specific route for vibrational relaxation.  The
key to our understanding of this specificity was the idea of taking an
{\it instantaneous} perspective:  of looking at each liquid
configuration and asking what the subsequent time evolution ought to be
for short time intervals.  The sharply varying character of the largest
magnitude intermolecular forces tells us that out of all of the solvent
molecules that could participate in the relaxation at a given instant,
only the very nearest solvent should play a significant role in
determining the high--frequency component of the solute--solvent
coupling.  This observation is central because it means that we only
need to ascertain the dynamics of the solute and the special solvent.
What is both more important and more unexpected though, are two
realizations stemming from the basic properties of liquids:  first,
when the liquid structure is such as to have the special solvent molecule
closer to the solute than to any other molecule in the solution (i.e., when
the solute and the special solvent molecule are a
mutual--nearest--neighbor--pair), then the pair motions themselves will
dynamically decouple from those of the rest of the system; second, that
these particular liquid--disorder--induced situations are precisely the
critical ones dominating the high--frequency relaxation dynamics.  For
the kinds of intermolecular distances and potentials typically found in
liquids, then, we see that the crucial dynamics is simply that of a
two--body scattering process.

Basic to all of this discussion is our assumption that the critical
portion of the time evolution takes place at short times.  In part,
this assumption is valid because of our concern about high
frequencies,~\cite{oxtoby77} but equally, it stems from the short (few
hundred fs) decay times characteristic of the force--autocorrelation
function which serves to determine the vibrational friction.  Of
course, this time scale, as well, has its origin in the same
fundamental ideas that we have been discussing.  The harsh repulsive
forces lead to a rapid decorrelation of the molecular trajectories and
the average over the structural disorder of the liquid washes out any
longer--time correlations that might remain.

The presence of this short time scale was important to us not only in
that it let us formulate our theory, but in its implications for the
choice between V--V and V--T mechanisms.  The fact that the dynamics
has a scattering flavor is certainly more reminiscent of the gas phase
than the solid state, but what we discovered was that for times this
short, the distinctions simply evaporate.  There is no discernable
difference between the high--frequency vibrational friction evaluated
as a scattering problem and the friction evaluated with
instantaneous--harmonic --- vibration--like --- bound--state motion.
As long as the coupling is sufficiently nonlinear, harmonics of the
fundamental vibration can evidently play the precise role occupied by
more impulsive kinematics.  The short answer to our basic question is
therefore that we can {\it always} think of solute as resonantly
transfering energy to vibrational motions of the solvent; to the
instantaneous normal modes for solute energies within the band of the
liquid and to overtones of those modes for higher energies.  This last
interpretation is obviously not unique, but it has the conceptual
advantage of allowing us to think of the progression from lower to
higher vibrational energies as a gradual evolution in the kinds of
dynamics the liquid has to bring to bear on the relaxation.

This continued success of harmonic perspectives on liquid dynamics
actually sends yet another message by highlighting one of the issues we
raised in the Introduction.  It is  hardly disputable that
understanding anharmonicity is vital to understanding liquid dynamics,
but the distinctions between the different flavors of anharmoncity have
significant consequences.  If we needed to be able to model the
detailed breakdown of the harmonic--mode picture of the underlying
dynamics in order to treat vibrational relaxation, we would have had a
daunting task in front of us.  Indeed, before we launched this work,
this worrisome prospect was a legitimate concern; over the long times
that it actually takes a molecule to rid itself of its vibrational
energy, individual instantaneous normal modes completely lose any
semblance of their original identities.~\cite{david98}  What we have
discovered, though, is that the primary issue in high-frequency
behavior is the nonlinearity of the coupling being driven by the liquid
dynamics, not the nonlinearity of the dynamics per se --- and that this
much more limited variety of anharmonicity is not all that difficult to
understand.

This same kind of distinction is actually a rather familiar one to
spectroscopists.  Early on in the history of infrared and Raman
spectra, distinctions were drawn between ``mechanical'' and
``electrical'' anharmonicities in the vibrational spectra of individual
molecules.~\cite{herzberg45}  More recently, however, the issue has
been the focus of considerable attention in the context of trying to
understanding fifth--order nonresonant Raman spectra of liquids as a
whole.~\cite{okumura97,tanimura93,okumura96,tokmakoff97,murry97,steffen,tominaga95,saito98}
Purely harmonic systems would give rigorously zero signal in these
experiments, so the very existence of a measurable signal is evidence
of anharmonicity.  But as Okumura and Tanimura were quick to point
out,~\cite{okumura97,okumura96} the same signals could arise equally
well from fundamental liquid--dynamical anharmonicities or from
nonlinearities in the polarizability dependence on normal modes
(violations of the Placzek approximation).~\cite{murry97}  Just as we
found here, the contribution of derivatives beyond the first for a
spectroscopic probe potential (whether it be a force on a bond or a
many--body polarizability) can be quite effective in extending the ways
a liquid can respond to a probe.  In fact, in perfect analogy to what
we saw here, the polarizability nonlinearities can apparently be a much
larger determinant of the spectra than the underlying liquid
anharmonicities.~\cite{nonlin-pol}

Curiously, the results of this paper seem as if they can also be taken
as supporting the independent--binary--collision model of vibrational
relaxation.  The centrality of solute--solvent pairs in the process
was, if anything, put on firmer microscopic ground by our results.
However, there are some key distinctions between IBC theory and what we
have presented here.  The most important of these is where the
equilibrium considerations stop and where the dynamical concerns take
over.  Within IBC theory the existence, the precise definition, and,
{\it a fortiori}, the rate of collisions all rely heavily on the
liquid--state setting.  Only the collision--induced vibrational
relaxation itself is thought to be independent of the
medium.~\cite{chesnoy88,davis72,oxtoby77,chesnoy86,paige90,fixman61,dardi88}
With our instantaneous--pair model, though, the sole role of the
many--body environment is to prescribe the {\it equilibrium}
distribution of special (mutual--nearest--neighbor) solute--solvent
pairs.  These pair distances form the initial configurations for the
dynamical calculation, but all of this subsequent dynamics, including
the distribution of initial pair velocities, are purely few--body in
character.  We should emphasize that these distinctions are not merely
technical.  The arbitrariness of IBC theory results from the difficult
task it has of solving enough of the nonequilibrium statistical
mechanics of the solution to extract a cleanly separable pair motion
from the remaining background.  Lacking a definitive solution, IBC
applications are compelled to adopt ad hoc definitions of collisions,
postulating such criteria as minimum collision
radii.~\cite{chesnoy86,paige90}  By contrast, the crisp separation with
instantaneous pair ideas between many--body equilibrium information and
few--body dynamics, not only removes any need for guesswork, it might
even allow for future quantum mechanical treatments of vibrational
relaxation based on the relative ease with which we can do two--body
quantum scattering calculations.~\cite{sun65}

There are some other directions the instantaneous pair theory should be
extended in as well.  All of the numerical examples explored here were
based on the rather limited possibilities presented by a diatomic
solute dissolved in an atomic solvent.  The easiest generalization of
this work to polyatomic solutes and solvents would probably have us
focus on mutual nearest neighbor pairs between individual solute sites
and solvent sites.  However the success of such an approach is far from
assured; sites within a molecule are always going to be strongly
correlated.  Thus in spite of the strongly binary character already
seen in INM vibrational friction spectra with molecular
solvents,~\cite{ladanyi98} a fully general version of
instantaneous--pair theory remains to be demonstrated.  The potential
applicability to understanding how solvents mediate the vibrational
energy relaxation of complex multi--mode polyatomics, in particular,
presents a natural extension we find especially intriguing.

\vspace{1cm}

{\bf Acknowledgements}.  It is our pleasure to acknowledge thoughful
discussions with Dr. Edwin David and Professor Branka Ladanyi.  We are
glad to acknowledge, as well, the thought--provoking comments of
Professor Bruce Berne, whose questions provided much of the impetus for
this work, and to thank the other participants of the 1997 Vibrational
Relaxation Symposium (held as part of that year's March Meeting of the
American Physical Society) for a number of useful conversations.  REL
gratefully acknowledges the receipt of a graduate student travel award
from the Chemical Physical division of the American Physical Society.
This work was supported by NSF grants CHE--9417546 and CHE--9625498.

\newpage

\appendix

\section{Analytical Calculations with Nonlinear INM Theory}

The velocity average needed in Eq. (\ref{g-super-n}) is a Boltzmann
average over the initial velocities $v_0 = {\dot q}(0)$
\begin{eqnarray} V_n(t)\ \equiv\ \left( {k_B T \over \mu_{uv}}
\right)^{-1}\ \left< q^{n-1}(t)\ {\dot q}(t)\ {\dot q}(0) \right>\ ,
\label{first-eq} \end{eqnarray} with the mode displacements $q(t)$
specified by Eq. (\ref{inm-dyn}).  The value of such integrals will
depend on the initial coordinate $r(0)$ as well as on time, (as
indicated by the notation in the text), but we have suppressed this
dependence here for notational simplicity.  For the purposes of this
appendix, brackets will always refer to averages over initial
velocity.

Evaluating expressions of this form is performed most easily by
observing that we can calculate the average of the mode displacement
raised to any power \begin{eqnarray} \left< \left[ \alpha_0\ q(t)
\right]^m \right>\ =\ \left[-i\ Y(t)\right]^m\ H_m[i\ X(t) ]\ ,
\label{m-th-power} \end{eqnarray} in terms of the Hermite polynomials
$H_m(x)$,~\cite{abram72} \begin{eqnarray} H_0(x)\ =\ 1,\ H_1(x)\ =
2x,\ H_2(x)\ =\ 4\,x^2\ -\ 2, \ldots, \nonumber \end{eqnarray} the
constants $\alpha_0$ and $t_0$ defined in Eq. (\ref{def-params}), and
the functions $X(t)$ and $Y(t)$ specified in Eq. (\ref{hermite-sum}).
Equation (\ref{m-th-power}) results quite naturally from the integral
relation~\cite{gradshteyn} \begin{eqnarray} \left( {a \over \sqrt{\pi}}
\right)\ \int_{-\infty}^\infty dx\ (x\ +\ z)^m\ e^{-(a x)^2}\ =\ (2 i
a)^{-m}\ H_m(i a z)\ . \end{eqnarray}

To have Eq. (\ref{m-th-power}) help us with Eq. (\ref{first-eq}),
though, we need to express the remaining factor in the average in terms
of the ``INM basis'', that is, in powers of $q(t)$.  After some
algebra, we find \begin{eqnarray} {\dot q}(t)\ {\dot q}(0)\ &=&\ 2
\left( {k_B T \over \mu_{uv} } \right)\ X^2(t)\ \left\{ Z(t)
[\alpha_0\ q(t)]^2\ + [\alpha_0\ q(t)]\ -\ 1 \right\}\ , \cr\cr
Z(t)\ &=&\ { \cos{\omega_0 t} \over ( 1 - \cos{\omega_0 t} )^2}\ .
\label{q-dot-prod} \end{eqnarray}  Substituting Eq. (\ref{q-dot-prod})
into Eq. (\ref{first-eq}), repeatedly making use of Eq.
(\ref{m-th-power}), and employing the Hermite polynomial recursion
relation~\cite{abram72} \begin{eqnarray} H_{n+1}(x)\ =\  2x\ H_n(x) -
2n\ H_{n-1}(x) , \nonumber \end{eqnarray} then leads us directly to the
expression for $V_n(t)$ given in Eq. (\ref{hermite-sum}).

To evaluate the multiphonon sum we need to sum these integrals to all
orders --- which we can actually do for the exponential model, Eq.
(\ref{exp-probe}).  For this particular model, the first two
contributions to the multiphonon sum $V(t)$ given in Eq.
(\ref{exp-sum}) are \begin{eqnarray} \left[ { (-\alpha)^0 \over 0! }
\right]\ V_1(t)\ &=&\ \cos{\omega_0 t} \cr\cr \left[ { (-\alpha)^1
\over 1! } \right]\ V_2(t)\ &=&\ \gamma_0 \left( \cos{2 \omega_0
t}\ -\ \cos{\omega_0 t} \right)\ , \nonumber \end{eqnarray} which makes
for a particularly simple form for the linear and linear-plus-quadratic
order terms in the vibrational relaxation case ($\gamma_0 = 1$)
\begin{eqnarray} &V&(t)^{linear}\ =\ \cos{\omega_0 t}\cr\cr
&V&(t)^{linear\ +\ quadratic}\ =\  \cos{2\omega_0 t}\ , \nonumber
\end{eqnarray} results that clearly emphasize the one-- and two--phonon
character of these contributions.  Higher order term do not preserve
this special structure, but we can still perform the full sum by taking
advantage of the generating function for Hermite
polynomials:~\cite{abram72} \begin{eqnarray} \sum_{n=0}^\infty \left(
{z^n \over n!} \right)\ H_n(x)\ =\ \exp{(2 x z\ -\ z^2)}\ . \nonumber
\end{eqnarray} Application of this formula and a simple variant lead to
the final result given in Eq. (\ref{summed-form}).  The same result
could also be obtained much more directly by simply perfoming the
velocity average in Eq. (\ref{general-G}).

\newpage

\begin{table}[htb]

\caption[Rates of Vibrational Population Relaxation for the
Tuckerman--Berne Model System]{Rates of Vibrational Population
Relaxation for the Tuckerman--Berne Model System$^a$}

\vspace{1cm}

\begin{tabular}{cccc} $\omega/2 \pi c\ (cm^{-1})^b$ & &
$1/T_1\ (ps^{-1})$ & \\ \hline & IP Theory$^c$ & MD$^d$ & linear INM
Theory$^e$ \\ \cline{2-4} 31 & 1.3 & 6.3 & 7.2 \\ 46 & 1.2 & 5.7 & 6.6
\\ 77 & 0.85 & 2.9 & 4.7 \\ 154 & 0.23 & 0.35 & 3.0 $\times 10^{-2}$
\\ 231 & 4.5 $\times 10^{-2}$ & 4.4 $\times 10^{-2}$ & 2.2 $\times
10^{-7}$ \\ 270 & 1.7 $\times 10^{-2}$ & 1.5 $\times 10^{-2}$ & 0$^f$
\end{tabular}

\vspace{1cm}

$^a$ Landau--Teller--theory calculations for a vibrating diatomic
molecule dissolved in a supercritical atomic liquid at reduced density
$\rho \sigma^3 = 1.05$ and reduced temperature $k_B T / \epsilon =
2.5$.  Numerical values are those appropriate for a solute in a Xe
solvent; to convert rates to those for an Ar solvent, both $\omega$ and
$1/T_1$ need to be multiplied by $\tau_{Xe}/\tau_{Ar} = 3.47 ps / 2.16
ps = 1.61$. \\ $^b$ Vibrational frequency of the isolated diatomic
solute. \\ $^c$ Predictions from the full instantaneous pair theory. \\
$^d$ Exact (Landau--Teller) calculations from molecular dynamics. \\
$^e$ Predictions from linearized instantaneous--normal--mode theory.
\\ $^f$ This rate is zero to within the numerical accuracy with which
we can determine the magnitude of the INM density of states.

\label{rel-times} \end{table}

\begin{thebibliography}{99}
\bibitem{renormed-w} The most accurate results are usually obtained by
taking $\omega_0$ to be the solution--phase value of the vibrational
frequency rather than the isolated--molecule value, technically, a
renormalized perturbation theory result.
\bibitem{delta-fric} In writing the friction in this fashion we have
neglected a term proportional to $\delta(\omega)$.
\bibitem{mnn-dists}  Typical liquid--state values for the temperature
fall between the triple point and the critical point,  $0.7 < k_B
T/\epsilon < 1.4$.  Typical mutual-nearest-neighbor distances $r_{ab}$
are actually on the order of $0.93 \sigma$ in a Lennard--Jones liquid
(see, for example, Ref. \onlinecite{larsen97}), which would imply an
even larger value of the intermolecular force than we used in our
estimate.  Indeed, the fact that this small a change in the
intermolecular distance leads to a value of the force 3.5 times larger
is itself a good illustration why both the coupling and dynamics are
dominated by the solvent atoms nearest to the solute atoms.
\bibitem{c-m-force} It is worth pointing out that in deriving the Eq.
(\ref{c-m-eqn}), we did not need to assume that the intermolecular
force was larger than the intramolecular force; the intramolecular
forces operating on the two solute sites are necessarily equal and
opposite and therefore make no net contribution to the center--of--mass
motion.
\bibitem{fwhm-i2}  For I$_2$ dissolved in Xe, $0.5 \AA = 0.1
\sigma_{uv}$, whereas for a dense Lennard--Jones liquid, the full width
at half maximum of the mutual--nearest--neighbor distribution is about
$0.09 \sigma$.
\bibitem{liq-analog}  The concept of a liquid--state analog to the
multiphonon processes seen in solids was also discussed recently by
Egorov and Berne (Ref. \onlinecite{quantum}).
\bibitem{anharm-stuff}  The analogous resummation for solid--state
multiphonon vibrational relaxation is well known.  See Refs.
\onlinecite{egorov95_2} and \onlinecite{nitzan75}.  However we believe
the results given here present microscopic evidence that high frequency
vibrational relaxation in liquids can be interpreted in a multiphonon
language.  Indeed, our INM results are somewhat different from the
expressions that appear in the solid--state literature, not so much
because of the quantum mechanical character of the literature
derivations (which give results identical in form to those from a
classical derivation), but because of the INM approach's explicit
inclusion of anharmonicity in the distribution of initial conditions.
\bibitem{lj-xe-ref} Our LJ parameters for xenon are identical to those
used in the simulations of Brown, Harris and Tully, Ref.
\onlinecite{brown88}.
\bibitem{init-vel}  The initial velocities selected, $v_0$, were
averaged as $(v_0,\ -v_0)$ pairs so as to minimize numerical noise.
\bibitem{one-dim}  In order to facilitate comparisons, the integrator
and the time step employed here were the same as that for the full
molecular dynamics simulation, velocity--Verlet and $0.001 \tau_{LJ}$,
respectively.  However since the instantaneous--pair theory requires
only a one--dimensional trajectory, it would have been computationally
trivial to use a higher order integrator and a far smaller time step
had we needed more reliable high--frequency information.
\bibitem{pair-stable}  Inasmuch as we were careful to use the same
molecular dynamics integrators and the same size time steps with the
pair--theory and with the full molecular dynamics calculation, the pair
theory's superior numerical stability in Figs. \ref{LJ-fric-fig} and
\ref{inm-fric-fig} is striking.  We attribute the difference to the
much higher level of velocity averaging possible with the pair theory.
\bibitem{alpha-ref}  The values of $\alpha_0$ turn out to be
surprisingly constant, suggesting that even a Lennard--Jones potential
looks remarkably exponential insofar as high--frequency dynamics goes.
Over the full width at half maximum of the mutual--nearest--neighbor
distribution for this thermodynamic state ($r/\sigma = 0.89$ --
$0.96$), the value of $\alpha_0\ \sigma = 17.2 \pm 0.3$ for the
Lennard--Jones potentials used here.
\bibitem{inertial} Parenthetically, were we to recompute the
nonlinearly--coupled/harmonic--dynamics curve with inertial (constant
force) dynamics (not shown) instead of harmonic dynamics, sum rules
would still generate the correct initial $t^2$ decay for us, but the
trajectories would soon wander off into regions of impossibly high
potential energy, leaving us with nonsensical correlation functions.
\end{thebibliography}
\end{document}